\documentclass[apsrev4-2,twocolumn,twoside,footinbib,superscriptaddress,floatfix,bibliography]{revtex4-2}
\usepackage{times}
\usepackage{float}
\usepackage{multirow}
\usepackage[dvipsnames]{xcolor}
\usepackage{amsmath,mathtools}
\usepackage{physics}
\usepackage[scr=boondox,  % heavily sloped
cal=esstix]   % slightly sloped
{mathalpha}
\usepackage{mathrsfs}
\usepackage{amsthm}
\usepackage{amssymb}
\usepackage{amsbsy}
\usepackage{wasysym}
\usepackage{enumitem}
\usepackage{wasysym}
\usepackage[english]{babel}
\usepackage[T1]{fontenc}
\usepackage[utf8]{inputenc} 
\usepackage{graphicx}
\usepackage{svg}
\usepackage[colorlinks,bookmarks=false,citecolor=blue,linkcolor=red,urlcolor=blue]{hyperref}
\usepackage{pstricks}
\usepackage{rotating}			       % creates rotated page (landscape) 
\usepackage{tabularx,hhline}	
\usepackage[caption=false]{subfig}	
\usepackage{appendix}
\usepackage{comment}
\usepackage{breakurl}

\usepackage{lipsum}
\makeatletter
\newcommand*{\balancecolsandclearpage}{%
  \close@column@grid
  \clearpage
  \twocolumngrid
}
\makeatother

\begin{document}
%%%%%%%%%%%%%%%%%%%%%%%%%%%%%%%%%%%%%

\let\subsectionautorefname\sectionautorefname
\let\subsubsectionautorefname\sectionautorefname
\def\chapterautorefname~#1\null{Chapter~#1\null}
\def\sectionautorefname~#1\null{Section~#1\null}
\def\figureautorefname~#1\null{Fig.~#1\null}
\def\tableautorefname~#1\null{Table~#1\null}
\def\equationautorefname~#1\null{Eq.~(#1)\null}

%%%%%%%%%%%%%%%%%%%%%%%%%%%%%%%%%%%%%
%\title{Taming entanglement at finite temperature: how does a ``classical'' paramagnet become a quantum spin liquid ?}
\title{Characterizing entanglement at finite temperature: how does a ``classical'' paramagnet become a quantum spin liquid ?}
%%%%%%%%%%%%%%%%%%%%%%%%%%%%%%%%%%%%%

\author{Snigdh Sabharwal}
\affiliation{Theory of Quantum Matter Unit, Okinawa Institute of Science and Technology Graduate University, Onna-son, Okinawa 904-0412, Japan}

\author{Matthias Gohlke}
\affiliation{Theory of Quantum Matter Unit, Okinawa Institute of Science and Technology Graduate University, Onna-son, Okinawa 904-0412, Japan}

\author{Paul Skrzypczyk}
\affiliation{H. H. Wills Physics Laboratory, University of Bristol, Tyndall Avenue, Bristol, BS8 1TL, UK.}

\author{Nic Shannon}
\affiliation{Theory of Quantum Matter Unit, Okinawa Institute of Science and Technology Graduate University, Onna-son, Okinawa 904-0412, Japan}

%%%%%%%%%%%%%%%%%%%%%%%%%%%%%%%%%%%%%
\date{\today}
%%%%%%%%%%%%%%%%%%%%%%%%%%%%%%%%%%%%%

%%%%%%%%%%%%%%%%%%%%%%%%%%%%%%%%%%%%%
\begin{abstract}
%%%%%%%%%%%%%%%%%%%%%%%%%%%%%%%%%%%%%

Quantum spin liquids (QSL) are phases of matter which are distinguished not by the 
symmetries they break, but rather by the patterns of entanglement within them. 
Although these entanglement properties have been widely discussed for ground states, 
the way in which QSL form at finite temperature remains an open question. 
Here we introduce a method of characterizing both the depth and spatial structure of 
entanglement, and use this to explore how patterns of entanglement form as temperature 
is reduced in two widely studied models of QSL, the Kitaev honeycomb model, and 
the spin-1/2 Heisenberg antiferromagnet on a Kagome lattice. 
These results enable us to evaluate both the temperature at which spins within the 
high-temperature paramagnet first become entangled, and the temperature at which 
the system first develops the structured, multipartite entanglement characteristic 
of its QSL ground state.

%%%%%%%%%%%%%%%%%%%%%%%%%%%%%%%%%%%%%
\end{abstract}
%%%%%%%%%%%%%%%%%%%%%%%%%%%%%%%%%%%%%

%%%%%%%%%%%%%%%%%%%%%%%%%%%%%%%%%%%%%
\maketitle
%%%%%%%%%%%%%%%%%%%%%%%%%%%%%%%%%%%%%

In 1935, as part of a broader debate about the interpretation of quantum mechanics, 
Schrödinger introduced 
the term {\it ``entanglement''} 
to describe the way in which mutually-exclusive experimental outcomes could be 
woven together in a single quantum mechanical 
state \cite{Schrodinger1935-Naturwissenschaften23,Schroedinger1935-MathProcCamPhilSoc31}.
Ninety years later, entanglement is no longer seen as a philosophical curiosity, 
but rather a technological resource \cite{Wootters1998}, with quantum advantage sought in the entanglement 
of scores of qubits \cite{Kim2023,Abanin2025}.
In the context of condensed matter physics, the most celebrated example of entanglement 
arises in quantum spin liquids (QSL) \cite{Lee2008,Balents2010,Savary2016,Broholm2020}, 
exotic magnetic phases where, at comparable temperatures, billions of spins can be entangled.
The sharpest distinctions between QSL, conventional magnetically--ordered phases, 
and states driven by disorder, are all expected to be found in their entanglement properties \cite{Wen2007,Amico2008,Laflorancie2016,Zeng2019,Broholm2020,Wu2019,Sabharwal2025,Shimokawa-arXiv.2505}, 
and significant results have also been found in the context of quantum codes~\cite{Dennis2002,Castelnovo2008,Hastings2011,Hart2018,Lu2020}.
None the less, little is known about how the entanglement characteristic of QSL behaves at 
finite temperature, or even whether it makes sense to discuss entanglement in candidate QSL materials, 
at the temperatures relevant to experiment.

%%%%%%%%%%%%%%%%%%%%%%%%%%%%%%%%%%%%%%
%  Fig. 1 - conceptual figure illustrating relevant cycles
%%%%%%%%%%%%%%%%%%%%%%%%%%%%%%%%%%%%%%

\begin{figure}[t]
    \centering
	   \subfloat[ Evolution of QSL out of high--temperature paramagnet \label{fig:2-step.scenario} ]{\includegraphics[width=0.9\columnwidth]{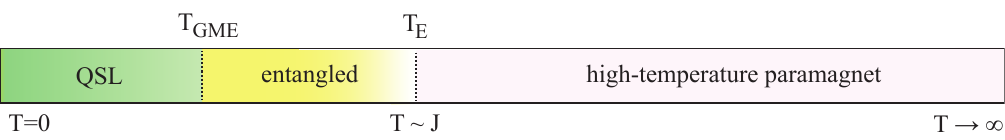}}
	\hspace{0.5cm}       
       \subfloat[ Kagome antiferromagnet \label{fig:kagome} ]{\includegraphics[width=0.9\columnwidth]{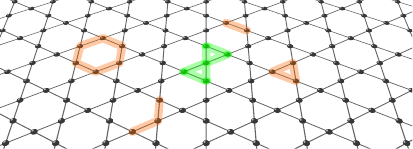}}\\
	\hspace{0.5cm}
	   \subfloat[ Kitaev honeycomb model \label{fig:honeycomb} ]{\includegraphics[width=0.9\columnwidth]{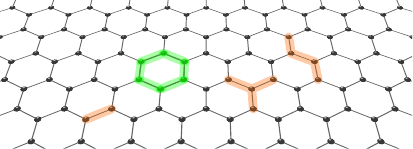}}
    \caption{
    Evolution of a quantum spin liquid (QSL) out of a high temperature 
    paramagnet.
    (a) Two--step scenario:
    on cooling, quantum effects enter at a temperature $T_{\sf E}$, 
    comparable with the interactions enforcing 
    local spin correlations.  
    Structured entanglement, characteristic of the QSL ground state, forms 
    at a lower temperature, $T_{\sf GME}$.
    This entanglement is multipartite in nature and, in the examples studied, 
    is confined to cycles of spins.     
    (b) For the Kagome antiferromagnet (KAF), the shortest relevant cycle is 
    the 6--link ``bow--tie'', highlighted in green.
    (c) For the Kitaev honeycomb model (KHM), the shortest relevant cycle is 
    a 6--link hexagonal ring.  
    }
    \label{fig:every.picture.tells.a.story}
\end{figure}

%%%%%%%%%%%%%%%%%%%%%%%%%%%%%%%%%%%%%

Answering these questions, even in the context of a theoretical model, poses a serious technical challenge.
The  most widely--used tool to date, entanglement entropy, 
is well--posed only for pure states at zero temperature \cite{Bennett1996,Kitaev2006-PRL96,Levin2006}.
\textcolor{black}{
The study of ``entanglement negativity", a proxy for the non-separability of a quantum state \cite{Zyczkowski1998,Eisert1999,Vidal2002}, 
has provided insight into the fate of topological entanglement at finite temperature \cite{Hart2018,Lu2020}, 
and the temperature-dependence of entanglement at quantum critical points \cite{Wang2025}.
A number of experimentally--accessible ``entanglement witnesses'' have been identified which are also applicable 
at finite temperature \cite{Guehne2009,Hauke2016,Scheie2021,Scheie2024,Laurell2024,Sabharwal2025,Shimokawa-arXiv.2505}.
However, each of these approaches is limited in the type of entanglement it can bear witness to, and 
evaluating more general measures of multipartite entanglement in mixed states, at finite temperature, involves 
optimization over all possible convex mixtures, and is an NP--hard problem \cite{Gurvits2003,Doherty2014}.
}
Fortunately, recent development of methods based on semi--definite programming (SDP) 
bring such questions into reach, making it possible to place precise, quantitative bounds on the 
entanglement encoded in the (reduced) density matrix for a given group of 
spins \cite{Jungnitsch2011,Hofmann2014,Skrzypczyk2023}. 
\textcolor{black}{
The value of these methods has been demonstrated in recent applications 
to the structured multipartite entanglement found in QSL 
ground states \cite{Lyu-arXiv.2505}, 
network aspects of entanglement \cite{Lyu-arXiv.2512}, 
and the impact of coupling to a QSL to a Bosonic bath \cite{Garcia-Gaitan-arXiv.2510}.
}
None the less, the fate of the entanglement which characterizes QSL 
at finite temperature remains an open question.

In this Letter we address the question of how the entanglement characteristic 
of a QSL is formed, as a frustrated spin system is cooled from a 
high-temperature paramagnet towards a QSL ground state.
To accomplish this, we introduce a method of characterizing 
the \textit{depth of entanglement}
~\cite{Sorensen2001,Guehne2005} present in an arbitrary subset 
of spins embedded within a larger system.
This is accomplished by means of the criterion of positive partial 
transpose (PPT) \cite{Peres1996,Horodecki1996}, 
evaluated within an SDP.  
We use this approach to explore how patterns of entanglement form in clusters 
of spins of different geometry as a function of temperature, considering two 
widely--studied models of QSL: the spin-1/2 Heisenberg antiferromagnet 
on a Kagome lattice, and the Kitaev honeycomb model, and calculating the necessary 
reduced density matrices using the method of thermal pure quantum states (TPQ).
In both cases we find that multipartite entanglement is restricted to cycles of spins, 
and to identify two characteristic temperatures: the temperature at which spins 
within the high-temperature paramagnet first become entangled, and the temperature 
at which the system first develops the structured, multipartite entanglement characteristic of 
its QSL ground state.
These results are summarised in Fig.~\ref{fig:every.picture.tells.a.story}

%%%%%%%%%%%%%%%%%%%%%%%%%%%%%%%%%%%%%%
% Paragraph 4
%%%%%%%%%%%%%%%%%%%%%%%%%%%%%%%%%%%%%%

The realization that magnetic insulators could host highly--entangled phases of matter, 
distinct from any conventional form of magnetic order, has almost as long a history 
as the idea of entanglement.
First explored in the context of one--dimensional spin chains~\cite{Hulthen1938}, 
the extension to higher dimension came through analogies with metals \cite{Pomeranchuck1941}, 
and resonating valence bonds \cite{Anderson1973}.
However the idea of a ``quantum liquid'' of spins didn't resonate widely until invoked 
to explain cuprate high--temperature superconductivity \cite{Anderson1987}. 
After decades of intensive study, QSL are now firmly established as part 
of the landscape of condensed matter physics \cite{Lee2008,Balents2010,Savary2016,Broholm2020},
with notable examples including the 
Kitaev honeycomb model (KHM) \cite{Kitaev2006-AnnPhys321,Baskaran2007,Knolle2015,Nasu2015,Gohlke2018,Hermanns2018,Rousochatzakis2019}, 
sought in ruthenates and iridates \cite{Jackeli2009,Matsuda-arXiv.2501}; 
the spin--1/2 Heisenberg antiferromagnet on a Kagome lattice (KAF) 
\cite{Sachdev1992,Chalker1992,Lecheminant1997,Hastings2000,Singh2007,Yan2011,Iqbal2011,Jiang2012,Liao2017,He2017,Laeuchli2018,Jiang2023}, 
discussed in the context of a wide range of materials \cite{Norman2016}; 
and quantum spin ice (QSI) \cite{Hermele2004,Benton2012,Lee2012}, 
actively pursued in rare earth pyrochlore oxides \cite{Sibille2018,Poree2025,Smith2025,Gao2025}.
None the less, because QSL are difficult to characterize, their unambiguous identification 
still presents many challenges, in both simulation and experiment.

%%%%%%%%%%%%%%%%%%%%%%%%%%%%%%%%%%%%%%
% Paragraph 5
%%%%%%%%%%%%%%%%%%%%%%%%%%%%%%%%%%%%%%

While conventional phases of matter are distinguished by order parameters which 
retain their meaning in a classical context, QSL are intrinsically quantum 
phases of matter, formed through the superposition of an extensive number of classical 
spin configurations, and capable of supporting fractionalized excitations with a topological character.
By construction, such states are entangled, and the structure of the 
entanglement within QSL is intimately linked to the structure of their 
excitations and the underlying lattice.
In the context of 2D QSL ground states, it has been widely argued that cycles play a special role in the 
entanglement properties \cite{Broholm2020,Wen2007,Zeng2019}.
More generally, the way in which different groups of spins contribute to entanglement will 
depend on the details of the QSL under consideration. 

%%%%%%%%%%%%%%%%%%%%%%%%%%%%%%%%%%%%%%
% Paragraph 6
%%%%%%%%%%%%%%%%%%%%%%%%%%%%%%%%%%%%%%

Testing these ideas requires a tool which is capable of resolving both the  
spatial structure of entanglement, and of distinguishing the mutlitpartite
entanglement characteristic of QSL from bipartite entanglement.
And for results to have bearing on experiment, this tool must also be 
applicable to the mixed states found at finite temperature. 
Calculations of entanglement entropy can be very revealing, but are 
limited to ground states \cite{Yao2010,Yan2011,Jiang2012}.    
Quantum Fisher Information (QFI) \cite{Braunstein1994}, which provides a bound 
on entanglement depth \cite{Hyllus2012,Toth2012}, does not suffer from 
this limitation \cite{Hauke2016}, and shows promise as a tool for distinguishing 
different quantum phases of matter through its temperature dependence
\cite{Shimokawa-arXiv.2505,Zhou-arXiv.2510}.  
None the less, QFI is limited by the fact that it is tied to a particular 
linear operator (commonly, the lattice Fourier transform of a spin operator \cite{Hauke2016}),  
which must be specified in advance, and is only sensitive 
to a limited set of entangled states.
A more universal witness is the Hamiltonian, which has been shown 
to provide information about the depth of entanglement  at finite temperature 
in spin chains and some 2D frustrated spin models \cite{Toth2005,Guehne2005,Guehne2006}.
However this approach is not easily adapted to the question 
of determining the spatial structure of entanglement in a QSL.

%%%%%%%%%%%%%%%%%%%%%%%%%%%%%%%%%%%%%%
% Paragraph 7
%%%%%%%%%%%%%%%%%%%%%%%%%%%%%%%%%%%%%%

Happily, while condensed--matter physicists were debating the existence of 
quantum spin liquids, the quantum information community were building 
new tools capable of characterizing entanglement in mixed states, 
see e.g.~\cite{Horodecki1996,Terhal2000,Guehne2009}.
A powerful paradigm is the detection and quantification of entanglement directly from the density matrix describing a group of qubits (spins), using tools of convex optimization \cite{Skrzypczyk2023}. 
In particular, by relaxing separability to positive partial transposition, 
the problem of quantifying entanglement can be expressed as a semidefinite 
program, which can be efficiently solved. 
Moreover, by exploiting the \textit{duality} of SDP's, this approach can also 
be used to generate entanglement witnesses which are specifically tailored 
to the system of interest. 
This overcomes the high rate of ``false negatives'' associated with using 
an ``off--the--shelf'' entanglement witness.
The resulting witnesses have two other very desirable qualities: firstly that 
it retains universal validity (i.e. it can be used to detect the entanglement 
of other density matrices), and secondly that it provides \textit{quantitative} 
information about entanglement for all states, by providing a lower bound on the 
``robustness'' of that entanglement \cite{Vidal1999,Steiner2003,Skrzypczyk2023}.

%%%%%%%%%%%%%%%%%%%%%%%%%%%%%%%%%%%%%%
% Paragraph 8
%%%%%%%%%%%%%%%%%%%%%%%%%%%%%%%%%%%%%%

In the context of QSL, an important step was the publication
of a recipe for quantifying multipartite entanglement, based on
an SDP \cite{Jungnitsch2011,Hofmann2014}.
Since this recipe applies to mixed states, it can be used to study 
the microscopic structure of entanglement, by applying it to 
reduced density matrices for clusters of spins within an extended lattice  \cite{Wang2025}. 
The power of this approach has already been demonstrated in the 
context of quantum critical points \cite{Wang2025}, 
and it was recently applied to 2D QSL at $T=0$, 
where it was found that multipartite entanglement was restricted 
to cycles of spins \cite{Lyu-arXiv.2505}.  
None the less, many questions remain open, not least the fate 
of the entanglement associated with these QSL ground states 
at finite temperature.
And to answer this question it would be helpful to have a tool 
capable of characterizing the temperature dependence of the 
depth of entanglement, i.e. the minimum number of qubits 
which \textit{must} be entangled for a given (mixed) state to exist.
In what follows, we introduce such a tool, and use it to explore 
how entanglement forms in two widely studied models, 
the spin--1/2 Heisenberg AF on a Kagome lattice, and the Kitaev honeycomb model, 
as they are cooled towards their QSL ground states.

%%%%%%%%%%%%%%%%%%%%%%%%%%%%%%%%%%%%%%
% Paragraph 8
%%%%%%%%%%%%%%%%%%%%%%%%%%%%%%%%%%%%%%

{\it SDP for entanglement depth}.   
%. 
We now introduce an SDP capable of placing a bound on {\it entanglement depth} \cite{Guehne2005}.  
The input for these calculations is the reduced density matrix $\rho_n(T)$ 
describing a cluster of $n$ qubits (spins) embedded within a larger lattice, 
at temperature~$T$.
Our goal is to determine whether or not it is possible to write this 
density matrix as a probabilistic mixture (convex sum) of states 
involving the entanglement of no more than $k-1$ qubits.
If $\rho_n(T)$ fails this test, then it must be entangled to (at least)
depth $k$.

%%%%%%%%%%%%%%%%%%%%%%%%%%%%%%%%%%%%%%
% Paragraph X
%%%%%%%%%%%%%%%%%%%%%%%%%%%%%%%%%%%%%%

We begin by enumerating $\mathscr{P}_{n, k-1}$, the set of  
of possible partitions $\pi$ of $n$ qubits to depth $k-1$, 
i.e. such that each subsystem $\mathcal{s}$ contains at most $k-1$ qubits.
For example, a possible partition of $n=5$ qubits 
$ABCDE$, to depth \mbox{$k-1 = 2$}, 
is \mbox{$\pi = AB|CD|E$}, in which case the subsystems 
$\mathcal{s}$ are $AB$, $CD$ or $E$. 
We then test whether it is possible to write the state as  
\begin{eqnarray}
    \rho_n(T) = \sum_{\pi \in \mathscr{P}_{n, k-1}} q_\pi \rho_\pi \; , \; 
    q_\pi \geq 0 \;, 
    \sum_{\pi \in \mathscr{P}_{n, k-1}} q_{\pi} = 1 \; , 
    \label{eq:convex.sum}
\end{eqnarray}
with the condition that 
\begin{align}
    \rho_\pi &= \sum_i p_i^\pi \bigotimes_{\mathcal{s} \in \pi} \rho_{\mathcal{s}}^{(i)} \; ,& p_i^\pi \geq 0 \; , 
    \quad \sum_i p_i^\pi =1 \; , 
    \label{eq:seprability.of.rho.pi}
\end{align}
i.e. that there is no entanglement between different subsystems $\mathcal{s}$.
Since enforcing the condition of strict separability, Eq.~(\ref{eq:seprability.of.rho.pi}), 
is computationally challenging, we relax this to a condition of positive 
partial transpose, a necessary consequence of separability \cite{Skrzypczyk2023}. 
More precisely, we require that $\rho_\pi^{T_{\mathcal{s}}}\succeq 0$ for 
each $\mathcal{s}$, where $T_{\mathcal{s}}$ denotes transposition (in a fixed, 
but arbitrary basis) of \textit{only} the subsystem $\mathcal{s}$. 
Any density matrix $\rho_n(T)$ which fails the above test
must be entangled to depth at least~$k$, and will be designated ``\textit{depth-$k$ entangled}''.  

%%%%%%%%%%%%%%%%%%%%%%%%%%%%%%%%%%%%%%
% Paragraph 9
%%%%%%%%%%%%%%%%%%%%%%%%%%%%%%%%%%%%%%

This test can be transformed into a quantitative 
measure of  entanglement at depth $k$, by determining the \textit{robustness} of that entanglement \cite{Vidal1999,Steiner2003}, i.e. how much noise must be mixed 
with $\rho_n(T)$ before it can be written in the form Eq.~(\ref{eq:convex.sum}).
This approach leads to a 
\textit{relative entropy} \cite{Datta2009} 
\begin{eqnarray}
\mathcal{S}_k [\rho_n(T)]
	= \ln \left(1 + \mathcal{R}_k [\rho_n(T)] \right) \; , 
\label{eq:depth.of.entanglement}
\end{eqnarray}
where $\mathcal{R}_k [\rho_n(T)]$ is a (generalized) robustness of entanglement
at depth $k$.
This can be determined through the SDP
\begin{align}
\label{eq:depth.of.entanglement-SDP}
	1 + \mathcal{R}_k [\rho_n(T)] = \min_{\{\tilde{\rho}_\pi\}} & \quad   \Tr \hspace{-1em}\sum_{\pi \in \mathscr{P}_{n,k-1}} \tilde{\rho}_\pi \\
	\text{s.t.}	 &   \quad	\sum_{\pi \in \mathscr{P}_{n,k-1}} \tilde{\rho}_\pi  \succeq \rho_n(T), \nonumber \\
    & \quad \tilde{\rho}_\pi \succeq 0 \quad \forall \pi \in \mathscr{P}_{n,k-1}, \nonumber \\
    & \quad \tilde{\rho}_\pi^{T_\mathcal{s}} \succeq 0 \quad \forall \mathcal{s} \in \pi, \forall \pi \in \mathscr{P}_{n,k-1}, \nonumber 
\end{align}
where 
$\tilde{\rho}_\pi$ are subnormalised states.
(Further details of this SDP, including its relationship with optimized 
entanglement witnesses, are given in End Matter).

%%%%%%%%%%%%%%%%%%%%%%%%%%%%%%%%%%%%%%
%% Fig. 2 - entanglement measures for Kagome AF
%%%%%%%%%%%%%%%%%%%%%%%%%%%%%%%%%%%%%%

\begin{figure}[t]
    \centering
    \includegraphics[width=1.0\linewidth]{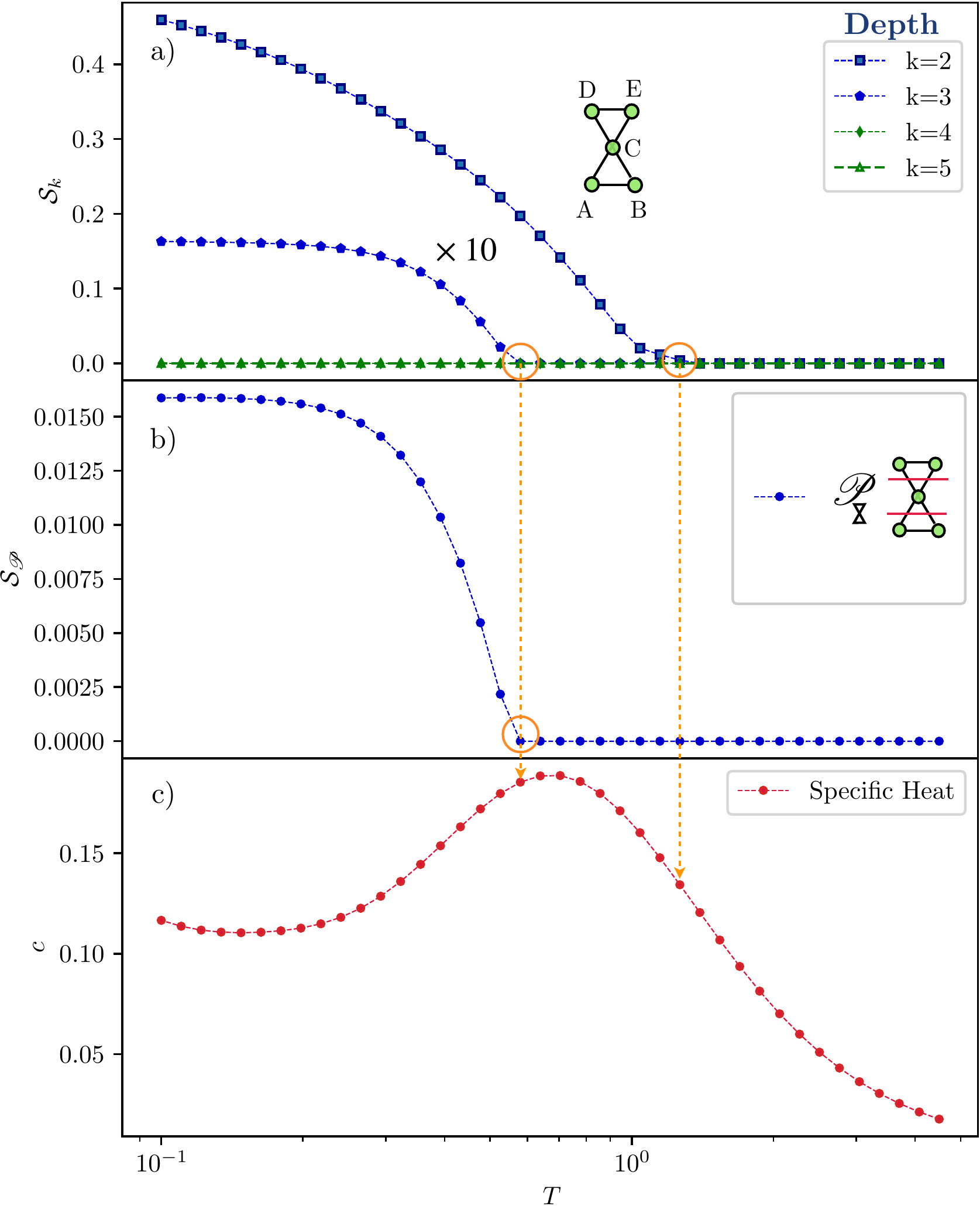}
    \caption{Evolution of entanglement as a function of temperature 
    in the spin-1/2 Heisenberg antiferromagnet on a Kagome lattice (KAF).
    (a)~Results for the depth of entanglement %(k-unproducability), 
    ${\mathcal S}_k  [\rho_n(T)]$ [Eq.~(\ref{eq:depth.of.entanglement})] 
    present  in the 6--link ``bow--tie'' cycle illustrated in Fig.~\ref{fig:kagome}.
    Entanglement on this cycle is first detected at a  
    temperature $T^{\bowtie}_{\sf E} = 1.26\ J$.  
    Multipartite entanglement of the type needed to support a QSL 
    is first detected at a temperature $T^{\bowtie}_{\sf GME} = 0.57\ J$.
    (b)~Corresponding results for genuine tripartite entanglement 
    ${\mathcal S}_{\sf GTE}  [\rho_n(T)]$, calculated through 
    Eq.~(\ref{eq:S.restricted.partitions}).
    An example of the type tripartition 
    % entering $\delta \mcathcal{P}_{\sf GTE}$ 
    considered is shown in the legend.
    (c)~Heat capacity $c(T)$, showing how 
 	changes in entanglement correlate with changes in thermodynamic properties.  
    Results are taken from thermal pure quantum state (TPQ) calculations 
    for a cluster of 24 spins, as described in the end matter.
    }
    \label{fig:results-for-Kagome}
\end{figure}

%%%%%%%%%%%%%%%%%%%%%%%%%%%%%%%%%%%%%%
% Paragraph 10
%%%%%%%%%%%%%%%%%%%%%%%%%%%%%%%%%%%%%%

Any value of 
\mbox{$\mathcal{S}_k [\rho_n(T)] > 0$} signals the presence of entanglement 
beyond subsystems of size $k-1$, and thus entanglement of depth $k$. 
If we consider $k = n$ (i.e.~maximal depth), 
\mbox{$\mathcal{S}_{k=n}  [\rho_n(T)]> 0$} implies that all subsystems are 
entangled, a situation more commonly known as 
\textit{genuine multipartite entanglement} (GME) \cite{Jungnitsch2011}.
More generally, information about the geometrical structure of entanglement 
can be obtained by restricting the set of partitions considered, vis. in \eqref{eq:depth.of.entanglement-SDP}
\begin{eqnarray}
\mathscr{P}_{n,k-1}  \to {\mathcal P}_\lambda:  \quad 
\mathcal{S}_k [\rho_n(T)] \to 
\mathcal{S}_\lambda [\rho_n(T)],
\label{eq:S.restricted.partitions}
\end{eqnarray}
where ${\mathcal P}_\lambda \subset \mathscr{P}_{n,k-1}$ is a subset of 
partitions tailored to the question under consideration.
We return to this point below, in the context of tripartite entanglement.

Finally, we note that, since we have relaxed separability to 
the PPT condition, our approach only detects entanglement 
associated with negative partial transposition. 
For this reason, \mbox{$\mathcal{S}_k  [\rho_n(T)]$} provides a 
lower bound on the amount of depth-$k$ entanglement in a system.
Where required, this bound can be progressively tightened, by strengthening 
the PPT condition, e.g. to shareability \cite{Doherty2014}, at the cost of 
a more complex SDP.  

%%%%%%%%%%%%%%%%%%%%%%%%%%%%%%%%%%%%%%
% Paragraph 11
%%%%%%%%%%%%%%%%%%%%%%%%%%%%%%%%%%%%%%

{\it  Results for Heisenberg antiferromagnet on a Kagome lattice}.   
We now turn to calculations for explicit models known (or in the case of the KAF, believed) 
to support QSL ground states. 
In Fig.~\ref{fig:results-for-Kagome}(a) we show results for 
the entanglement depth 
of the KAF as a function of temperature, 
evaluated on the ``bow--tie'' cycle shown in green in Fig.~\ref{fig:kagome}.
We find that density matrix on this cycle is nonseparable
for all temperatures \mbox{$T < T^{\bowtie}_{\sf E} = 1.26\ J$}.
Because of the specific structure of this cycle, entanglement 
at depth \mbox{$k=3$} acts as witness of tripartite entanglement, i.e.~GME.  
This is detected at a temperature \mbox{$T^{\bowtie}_{\sf GME} = 0.57\ J$}, 
just below a broad peak in heat capacity [Fig.~\ref{fig:results-for-Kagome}(c)].
No further thresholds for depth $k$ are detected for this cycle.
We have confirmed that this measure of GME vanishes at finite temperature 
on all other cycles and tree--like graphs containing up to 6 spins.
In contrast, bipartite entanglement at depth \mbox{$k=2$} can be detected on 
a number of graphs, including the triangular cycle shown 
in Fig.~\ref{fig:kagome}, and a decorated version of the bow--tie 
cycle with two additional spins (not shown).

%%%%%%%%%%%%%%%%%%%%%%%%%%%%%%%%%%%%%%
% Paragraph 12
%%%%%%%%%%%%%%%%%%%%%%%%%%%%%%%%%%%%%%

The KAF was one of the cases considered in 
Ref.~\cite{Lyu-arXiv.2505}, where Lanczos diagonalization 
was used to calculate ground--state properties for a cluster of 36 spins.
It was found that GME, studied as tripartite entanglement through a mapping 
onto three distinct ``parties'' of qubits~\cite{Jungnitsch2011}, and quantified 
through entanglement negativity~\cite{Hofmann2014}, was finite for the bow--tie cycle
discussed above, and vanished on all other graphs of up to 6 spins.   
We have repeated these calculations, finding identical results 
at \mbox{$T=0$}, and extended the analysis to finite temperature, 
by adapting the SDP for entanglement depth.
In Fig.~\ref{fig:results-for-Kagome}(b) we show results for 
tripartite entanglement, $\mathcal{S}_{\sf GTE} [\rho_n(T)]$, 
calculated through Eq.~(\ref{eq:S.restricted.partitions}), where 
${\mathcal P}_{\sf GTE}$ was matched to the partitions 
considered in \cite{Lyu-arXiv.2505} (see End Matter).  
Results %for triparite entanglement 
are numerically identical to the GME detected by 
$\mathcal{S}_{k=3} [\rho_n(T)]$~[Fig.~\ref{fig:results-for-Kagome}(a)].

%%%%%%%%%%%%%%%%%%%%%%%%%%%%%%%%%%%%%%
% Paragraph 13
%%%%%%%%%%%%%%%%%%%%%%%%%%%%%%%%%%%%%%

\begin{figure}[t]
    \centering
    \includegraphics[width=1.0\linewidth]{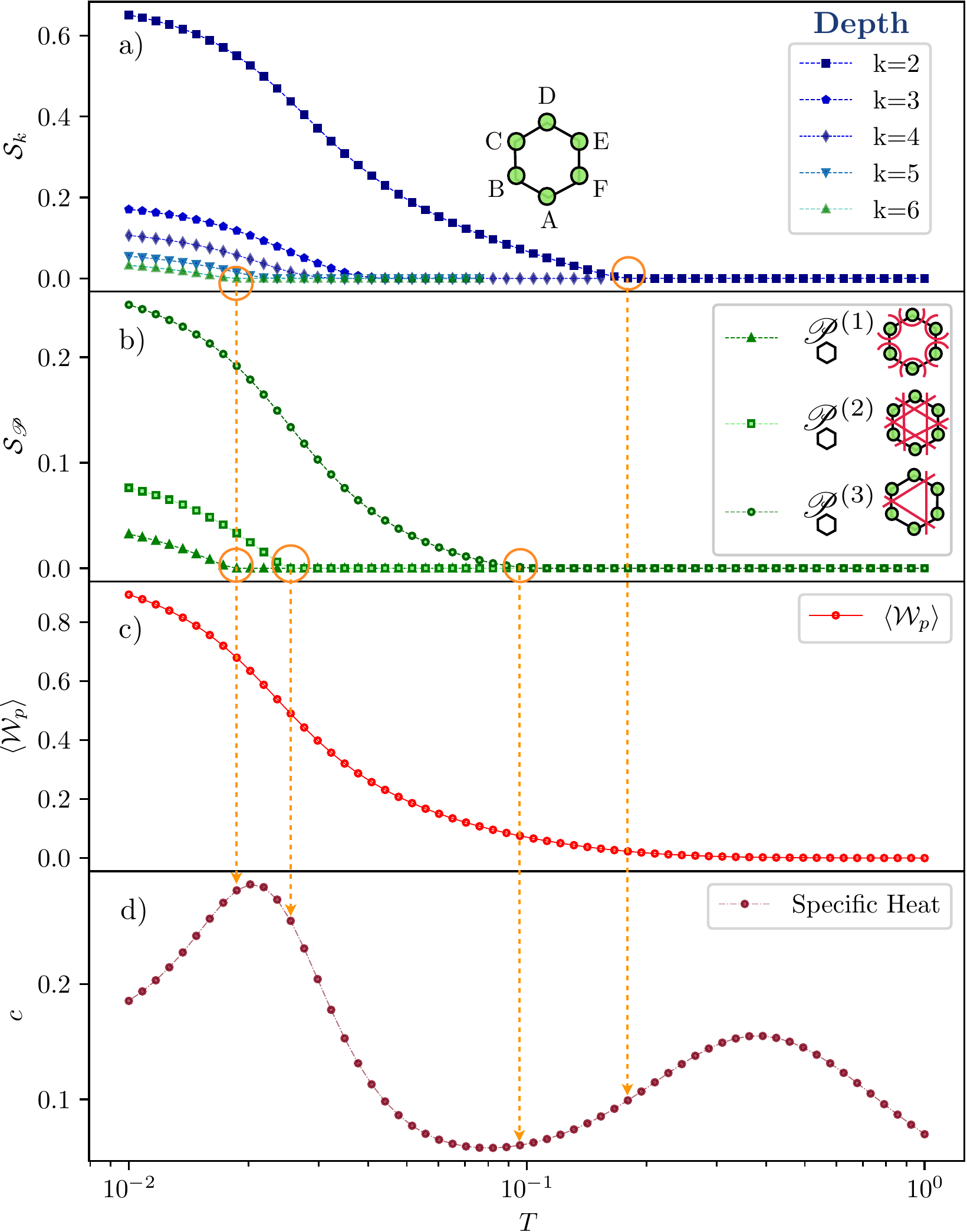}
    \caption{Evolution of entanglement as a function of temperature 
    in the Kitaev model on a honeycomb lattice (KHM).
    (a)~Results for the depth of entanglement, 
    ${\mathcal S}_k [\rho_n(T)]$ [Eq.~(\ref{eq:depth.of.entanglement})] 
    present in the six--link hexagonal cycle illustrated in Fig.~\ref{fig:honeycomb}.
    Entanglement on this cycle is first detected at a temperature  
    $T^{\hexagon}_{\sf E} = 0.18\ J$.   
    Genuine multipartite entanglement (GME) is detected 
    at a lower temperature $T^{\hexagon}_{\sf GME} =  0.018\ J$.
    (b)~Corresponding results for the entanglement ${\mathcal S}_\lambda [ \rho_{n}(T)]$, 
    associated with different geometrical partitions 
    \mbox{$ \mathcal{P}_\lambda = \mathscr{P}^{(m)}_{\hexagon}$} 
    [Eq.~(\ref{eq:S.restricted.partitions}), End Matter].
    Tripartite entanglement is detected via $\mathscr{P}^{(3)}_{\hexagon}$ 
    at a temperature $T^{\hexagon}_{\sf GTE} =0.096 \ J$.
    (c)~Expectation value of plaquette operator $\langle \mathcal{W}_p \rangle$ 
    [Eq.~(\ref{eq:W_p})] associated with $\mathbb{Z}_2$ gauge flux.
    (d)~Heat capacity $c(T)$~[Eq.~\ref{eq:C.of.T}], showing two--peak 
    structure characteristic of the Kitaev spin liquid.   
    Results are taken from thermal pure quantum state (TPQ) calculations 
    for a cluster of 24 spins, as described in End Matter.
    }
    \label{fig:results-for-Kitaev-model}
\end{figure}

{\it  Results for Kitaev honeycomb model}.   
We next consider how entanglement evolves at finite temperatures in 
KHM, focusing on the hexagonal cycle shown in Fig.~\ref{fig:honeycomb}.
In Fig.~\ref{fig:results-for-Kitaev-model}(a) we present results for 
depth of entanglement, quantified through 
${\mathcal S}_k[\rho_n(T)]$~[Eq.~(\ref{eq:depth.of.entanglement})].
We find that entanglement at depth $k=2$ is first detected 
at a temperature \mbox{$T^{\hexagon}_{\sf E} = 0.18\ J$} 
[Fig.~\ref{fig:results-for-Kitaev-model}(a)]. 
A cascade of bounds on depth for $k=3,4,5$ are found at lower temperature,
with GME revealed by $\mathcal{S}_{k=6} [\rho_n(T)]$ at 
\mbox{$T^{\hexagon}_{\sf GME} =  0.018\ J$}. 

%%%%%%%%%%%%%%%%%%%%%%%%%%%%%%%%%%%%%%
% Paragraph 14
%%%%%%%%%%%%%%%%%%%%%%%%%%%%%%%%%%%%%%

Identifying the geometrical structure of multipartite entanglement 
within the KHM proves to be a more delicate question.
In Fig.~\ref{fig:results-for-Kitaev-model}(b) we show results 
for the entanglement $\mathcal{S}_\lambda [\rho_n(T)]$ 
associated with different possible partitions 
$\delta{\mathcal P}_\lambda$, calculated through 
Eq.~(\ref{eq:S.restricted.partitions}) [see End Matter].
If we consider the strongest version of GME, with each spin considered 
separately, \mbox{${\mathcal P}_\lambda = {\mathscr P}_{\hexagon}^{(1)}$}, 
we recover \mbox{$T^{\hexagon}_{\sf GME} =  0.018\ J$}, as previously found through 
$\mathcal{S}_{k=6} [\rho_n(T)]$ [Fig.~\ref{fig:results-for-Kitaev-model}(a)].
The ground state of the KHM was another of the cases considered 
in \cite{Lyu-arXiv.2505}, where a different approach was taken, 
with the hexagonal cycle partitioned into three equal subsystems.
Following this prescription,  
\mbox{${\mathcal P}_\lambda = {\mathscr P}_{\hexagon}^{(3)}$},  
we find \mbox{$T^{\hexagon}_{\sf GTE} = 0.096 \ J$}.
However there is no unique way of dividing up the hexagon into three equal 
pieces, and if we include the symmetric nearest-neighbour alternative on a 
equal footing, ${\mathcal P}_\lambda ={\mathscr P}_{\hexagon}^{(2)}$, 
we find a significantly lower threshold \mbox{$T_{\hexagon}^{(2)} = 0.025\ J$}.
In the absence of a physical argument linking the QSL ground state 
to specifically tripartite entanglement, we make the conservative 
choice of associating GME with 
${\mathcal P}_\lambda = {\mathscr P}_{\hexagon}^{(1)}$, 
where agreement is found between 
${\mathcal S}_\lambda [\rho_n(T)]$ and ${\mathcal S}_{k=n} [\rho_n(T)]$.   
We have also confirmed that GME, regardless of definition, vanishes 
at finite temperature on all other cycles and tree--like graphs 
containing up to 6 spins.

%%%%%%%%%%%%%%%%%%%%%%%%%%%%%%%%%%%%%%
%% Fig. 3 - entanglement measures for Kitaev model on a honeycomb lattice
%%%%%%%%%%%%%%%%%%%%%%%%%%%%%%%%%%%%%%

%%%%%%%%%%%%%%%%%%%%%%%%%%%%%%%%%%%%%%
% Paragraph 15
%%%%%%%%%%%%%%%%%%%%%%%%%%%%%%%%%%%%%%

{\it General discussion of results:}
These results suggest a relatively simple qualitative picture of how entanglement 
forms as a high--temperature paramagnet is cooled towards a QSL ground state.   
Quantum fluctuations 
emerge once spins are sufficiently correlated to support them, generating
entanglement, and so breaking the separability of the state at a temperarture 
\mbox{$T_{\sf E} \lesssim J$}. 
This change will generally be accompanied by a peak in heat capacity, 
coming from the entropy associated with short--range classical spin 
correlations (e.g, a local constraint such as the ``ice rule'' in quantum spin ice \cite{Banerjee2008,Shannon2012,Benton2012,Kato2015,Huang2018}).
At lower temperatures, quantum fluctuations act coherently, as excitations 
about a well defined ``vacuum'' (QSL ground state), giving rise to multipartite 
entanglement on cycles of spins for a temperature \mbox{$T_{\sf GME} \ll J$}. 
A second, low--temperature peak in heat capacity will typically accompany the 
onset of the GME, reflecting further changes in entropy coming from quantum correlations.

%%%%%%%%%%%%%%%%%%%%%%%%%%%%%%%%%%%%%%
% Paragraph 16
%%%%%%%%%%%%%%%%%%%%%%%%%%%%%%%%%%%%%%

This simple scenario chimes with what is already known about the properties 
of the Kitaev honeycomb model at finite temperature, where comparison can be 
made with quantum Monte Carlo  \cite{Nasu2015}, TPQ \cite{Yamaji2016}, 
finite--temperature Lanczos  \cite{Rousochatzakis2019}
and tensor--network calculations \cite{Li2020,Gohlke2023}.
These calculations provide evidence for a two-step evolution of the QSL out of the 
high--temperature paramagnet, with the emergence of itinerant Majorana 
fermions correlating with a broad peak in heat capacity at \mbox{$T \approx 0.5\ J$}, 
and the onset of a coherent $\mathbb{Z}_2$ flux state [$\langle \mathcal{W}_{p} \rangle \to 1$]  [Eq.~(\ref{eq:W_p})], 
being signaled by a sharper peak in heat capacity at \mbox{$T  \approx 0.02\ J$}.
Unlike the coherent fluctuations associated with GME, the entanglement signatures of 
itinerant Majorana fermions in the intermediate temperature regime \mbox{$0.02\ J \lesssim T \lesssim 0.5\ J$}
of the Kitaev model, need not be confined to closed cycles of spins.
Consistent with this, we find that finite bipartite entanglement can 
also be detected in a decorated version of the hexagonal cycle 
containing 7 spins, 
at a slightly higher temperature \mbox{$T^{(7)}_{\sf E} \sim 0.2\ J$}. 
We aniticpate that, were it possible to solve for entanglement on 
arbitrarily large graphs, $T_{\sf E}$ would migrate to a temperature 
comparable with the upper peak in $c(T)$.
A more rigorous investigaton of this, and related finite--size effects, 
lies outside the scope of this manuscript, and is reserved for future work.
None the less, the present calculations for $N=24$ spins establish a clear correlation
between the rise in GME and the rise in $\langle \mathcal{W}_p \rangle$, 
with the onset of the most conservative measure of GME, ${\mathcal S}_{k=n} [\rho_n(T)]$,  
correlating well with the lower peak in $c(T)$.   

%%%%%%%%%%%%%%%%%%%%%%%%%%%%%%%%%%%%%%
% Paragraph X
%%%%%%%%%%%%%%%%%%%%%%%%%%%%%%%%%%%%%%

Much less is known about the finite--temperature properties of the KAF, 
with most studies focusing on the possibility 
of a QSL ground state \cite{Sachdev1992,Chalker1992,Lecheminant1997,Hastings2000,Singh2007,Yan2011,Iqbal2011,Jiang2012,Liao2017,He2017,Laeuchli2018,Jiang2023,Lyu-arXiv.2505}.
None the less, our results are consistent with what is known 
from TPQ calculations of thermodynamic properties for clusters 
of up to 36 spins~\cite{Sugiura2013,Shimokawa2016}, over
the range of temperatures considered in Fig.~\ref{fig:results-for-Kitaev-model}.
It is also possible to obtain an independent estimate of $T_{\sf E}$  
by comparing the energy $E(T)$ found in TPQ calculations  
with the lowest energy achievable through a separable state, 
$E^{\sf KAF}_{\sf sep}/N = - J/4$, following \cite{Dowling2004,Toth2005}.
For the KAF leads to a bound 
\mbox{$T_{\sf E} %= T^{\bowtie}_{\sf E} 
= 1.26\ J$}, 
identical to that found from the SDP for depth $k=2$ on a bow--tie cycle.
Applying the same technique to the KHM, 
where \mbox{$E^{\sf KHM}_{\sf sep}/N = -J/8$} \cite{Baskaran2008}, 
we find \mbox{$T_{\sf E} = 0.57\ J$}.
This lies a little above the upper peak in $c(T)$, consistent with 
general expectations.

%%%%%%%%%%%%%%%%%%%%%%%%%%%%%%%%%%%%%%
% Paragraph 19
%%%%%%%%%%%%%%%%%%%%%%%%%%%%%%%%%%%%%%

When it comes to experiments on quantum magnets, the main implications 
of these results are to provide a conceptual framework for discussing 
at what temperature a material becomes a QSL.
The weakest criterion is $T_{\sf E}$, the temperature below which bipartite
entanglement is present.    
This is expected to correlate with a wide range of interesting behaviour, including 
the emergence of fractionalized quasi--particles, seen e.g. as a continuum in 
inelastic scattering.
These considerations apply equally to fractionalized quasiparticles observed in 
proximate QSL \cite{Lake2013,Banerjee2016,Yamaji2016,Gohlke2017,Scheie2024}.
A much stricter criterion is $T_{\sf GME}$, the temperature at which entanglement 
starts to resemble a QSL ground state. 
The presence of GME, at least locally, is a necessary condition for those 
excitations which involve coherent fluctuations of an emergent gauge field, 
such as the photons of QSI \cite{Hermele2004,Benton2012}, or ${\mathbb Z}_2$  
vortices associated in the Kitaev model \cite{Kitaev2006-AnnPhys321}.  
And from this point of view it is interesting that GME within short cycles 
``turns on'' at a clearly--defined temperature for 
both of the QSL studied [Fig.~\ref{fig:results-for-Kagome}(a,b), 
Fig.~\ref{fig:results-for-Kitaev-model}(b)], behavior reminiscent of an order parameter.
No sharp anomaly is observed in $c(T)$ at these temperatures, but broad peaks are 
found at similar scales.  
And in the case of the KHM it is the most rigorous criterion for GME, ${\mathcal S}_{k=n} [\rho_n(T)]$ 
which best matches the low--temperature peak in $c(T)$, 
\textcolor{black}{
associated with entry into the flux-free state.  
It follows that in QSLs like the KHM which have a single, emergent, low energy scale, 
the presence of this peak may serve as a useful proxy for GME.
}

%'And in the case of the KHM it is the most rigorous criterion for GME, ..., which best matches the low-temperature peak in c(T) typically associated with aproaching the flux-free state <W_p> -> 0. Hence, for a pure Kitaev system, this peak serves as a proxy for GME. It remains an open question, whether such a thermodynamic proxy for GME can be identified in other QSL candidate systems.'

%%%%%%%%%%%%%%%%%%%%%%%%%%%%%%%%%%%%%%
% Paragraph 20
%%%%%%%%%%%%%%%%%%%%%%%%%%%%%%%%%%%%%%

{\it Conclusions and outlook: }
Quantum spin liquids are quantum phases of matter distinguished by the 
patterns written in their entanglement.
The patterns of entanglment found in QSL ground states are increasingly  
well understood, but the fate of this entanglement at finite temperature 
remains an open question. 
In this Letter, we have introduced a measure of the depth of entanglement 
found in a general mixed state, and used it to explore the way in which entanglement 
forms as a system is cooled from a high--temperature paramagnet towards a QSL ground state.
Considering both the Kagome lattice AF [Fig.~\ref{fig:results-for-Kagome}], and 
the Kitaev honeycomb model [Fig.~\ref{fig:results-for-Kitaev-model}], we have 
established a lower bound on the temperature at which spins first become entangled, 
and an upper bound on the temperature at which they exhibit the 
genuine multipartite entanglement needed to support a coherent QSL ground state.
We find that GME at finite temperature is limited to cycles of spins, 
as previously established for QSL ground states.  
These results provide a conceptual framework for understanding quantum 
spin liquids at finite temperature, and provide a starting point for interpreting the 
temperature scales over which a QSL can sensibly be discussed in experiment.

%%%%%%%%%%%%%%%%%%%%%%%%%%%%%%%%%%%%%%
% Paragraph 21
%%%%%%%%%%%%%%%%%%%%%%%%%%%%%%%%%%%%%%

The ability to characterise entanglement, and in particular the depth of 
entanglement, at finite temperature, also opens a new perspective on the 
question of quantum--classical correspondence: 
indeed this Letter provides two worked examples of how classical
separable states at high temperature evolve into an intrinsically quantum 
phase at low temperature.
While it is too early to speculate about the broader generality of the results, 
the tools used can be applied to any problem where a density matrix
can be calculated in simulation, or reconstructed from measurements 
made on a quantum simulator.   
The characterization of entanglement at finite temperature therefore looks 
certain to have an exciting future.
\\

%%%%%%%%%%%%%%%%%%%%%%%%%%%%%%%%%%%%%%
{\it Acknowledgements:}. 
%%%%%%%%%%%%%%%%%%%%%%%%%%%%%%%%%%%%%%
%
The authors are pleased to acknowledge helpful conversations with 
Jiahui Bao, Robert Joynt, Yoshi Kamiya, Liuke Lyu, Sandu Popescu, and Tokuro Shimokawa.
SS is grateful for the hospitality of the University of Bristol, where a part 
of this work was carried out.
This work was supported by the Theory of Quantum Matter Unit, OIST, 
and by CIFAR (through CIFAR Azrieli Global Scholarship funds).
We acknowledge the use of computational resources of the Scientific Computing section 
of the Research Support Division at the Okinawa Institute of Science and Technology 
Graduate University (OIST).

%%%%%%%%%%%%%%%%%%%%%%%%%%%%%%%%%%%%%%
% {\it Author contributions:} 
%%%%%%%%%%%%%%%%%%%%%%%%%%%%%%%%%%%%%%
%
%SS concieved of the SDP used to determine entanglement depth, 
%and carried out all calculations.
%
%MG provided input on the Kitaev model. 
%
%PS provided input on SDP's and entanglement measures. 
%
%NS provided input on QSL, and prepared the first draft of the manuscript.
%
%All authors contributed to the planning of the project, interpretation of results, 
%and preparation of the final version of the manuscript.

%%%%%%%%%%%%%%%%%%%%%%%%%%%%%%%%%%%%%%
 \bibliography{short_paper}

@misc{Lyu-arXiv.2512,
      title={Network-Irreducible Multiparty Entanglement in Quantum Matter}, 
      author={Liuke Lyu and Pedro Lauand and William Witczak-Krempa},
      year={2025},
      eprint={2512.11118},
      archivePrefix={arXiv},
      primaryClass={quant-ph},
      url={https://arxiv.org/abs/2512.11118}, 
}

@misc{Snigdhgit2025,
  author = {Sabharwal, Snigsh},
  title = {DepthPhi},
  year = {2025},
  publisher = {GitHub},
  journal = {GitHub repository},
  howpublished = {\url{https://github.com/seuc/DepthPhi}}
}

@misc{Garcia-Gaitan-arXiv.2510,
      title={Fate of entanglement in open quantum spin liquid: Time evolution of its genuine multipartite negativity upon sudden coupling to a dissipative bosonic environment}, 
      author={Federico Garcia-Gaitan and Branislav K. Nikolic},
      year={2025},
      eprint={2510.02256},
      archivePrefix={arXiv},
      primaryClass={cond-mat.str-el},
      url={https://arxiv.org/abs/2510.02256}, 
}

@article{Eisert1999,
author = {Jens Eisert and Martin B. Plenio},
title = {A comparison of entanglement measures},
journal = {Journal of Modern Optics},
volume = {46},
number = {1},
pages = {145--154},
year = {1999},
publisher = {Taylor \& Francis},
doi = {10.1080/09500349908231260},
URL = { 
https://www.tandfonline.com/doi/abs/10.1080/09500349908231260},
eprint = {https://www.tandfonline.com/doi/pdf/10.1080/09500349908231260}
}

@article{Vidal2002,
  title = {Computable measure of entanglement},
  author = {Vidal, G. and Werner, R. F.},
  journal = {Phys. Rev. A},
  volume = {65},
  issue = {3},
  pages = {032314},
  numpages = {11},
  year = {2002},
  month = {Feb},
  publisher = {American Physical Society},
  doi = {10.1103/PhysRevA.65.032314},
  url = {https://link.aps.org/doi/10.1103/PhysRevA.65.032314}
}

@article{Zyczkowski1998,
  title = {Volume of the set of separable states},
  author = {\ifmmode \dot{Z}\else \.{Z}\fi{}yczkowski, Karol and Horodecki, Pawe\l{} and Sanpera, Anna and Lewenstein, Maciej},
  journal = {Phys. Rev. A},
  volume = {58},
  issue = {2},
  pages = {883--892},
  numpages = {0},
  year = {1998},
  month = {Aug},
  publisher = {American Physical Society},
  doi = {10.1103/PhysRevA.58.883},
  url = {https://link.aps.org/doi/10.1103/PhysRevA.58.883}
}

@article{Amico2008,
	author = {Amico, Luigi and Fazio, Rosario and Osterloh, Andreas and Vedral, Vlatko},
	doi = {10.1103/RevModPhys.80.517},
	issue = {2},
	journal = {Rev. Mod. Phys.},
	month = {May},
	numpages = {0},
	pages = {517--576},
	publisher = {American Physical Society},
	title = {Entanglement in many-body systems},
	url = {https://link.aps.org/doi/10.1103/RevModPhys.80.517},
	volume = {80},
	year = {2008}}

@article{Dowling2004,
	author = {Dowling, Mark R. and Doherty, Andrew C. and Bartlett, Stephen D.},
	doi = {10.1103/PhysRevA.70.062113},
	issue = {6},
	journal = {Phys. Rev. A},
	month = {Dec},
	numpages = {15},
	pages = {062113},
	publisher = {American Physical Society},
	title = {Energy as an entanglement witness for quantum many-body systems},
	url = {https://link.aps.org/doi/10.1103/PhysRevA.70.062113},
	volume = {70},
	year = {2004}}

@article{Hart2018,
	author = {Hart, O. and Castelnovo, C.},
	doi = {10.1103/PhysRevB.97.144410},
	issue = {14},
	journal = {Phys. Rev. B},
	month = {Apr},
	numpages = {15},
	pages = {144410},
	publisher = {American Physical Society},
	title = {Entanglement negativity and sudden death in the toric code at finite temperature},
	url = {https://link.aps.org/doi/10.1103/PhysRevB.97.144410},
	volume = {97},
	year = {2018}}

@article{Dennis2002,
	author = {Dennis, Eric and Kitaev, Alexei and Landahl, Andrew and Preskill, John},
	doi = {10.1063/1.1499754},
	issn = {0022-2488},
	journal = {Journal of Mathematical Physics},
	month = {09},
	number = {9},
	pages = {4452-4505},
	title = {Topological quantum memory},
	url = {https://doi.org/10.1063/1.1499754},
	volume = {43},
	year = {2002}}

@article{Castelnovo2008,
	author = {Castelnovo, Claudio and Chamon, Claudio},
	doi = {10.1103/PhysRevB.78.155120},
	issue = {15},
	journal = {Phys. Rev. B},
	month = {Oct},
	numpages = {26},
	pages = {155120},
	publisher = {American Physical Society},
	title = {Topological order in a three-dimensional toric code at finite temperature},
	url = {https://link.aps.org/doi/10.1103/PhysRevB.78.155120},
	volume = {78},
	year = {2008}}

@article{Lu2020,
	author = {Lu, Tsung-Cheng and Hsieh, Timothy H. and Grover, Tarun},
	doi = {10.1103/PhysRevLett.125.116801},
	issue = {11},
	journal = {Phys. Rev. Lett.},
	month = {Sep},
	numpages = {6},
	pages = {116801},
	publisher = {American Physical Society},
	title = {Detecting Topological Order at Finite Temperature Using Entanglement Negativity},
	url = {https://link.aps.org/doi/10.1103/PhysRevLett.125.116801},
	volume = {125},
	year = {2020}}

@article{Hastings2011,
	author = {Hastings, Matthew B.},
	doi = {10.1103/PhysRevLett.107.210501},
	issue = {21},
	journal = {Phys. Rev. Lett.},
	month = {Nov},
	numpages = {5},
	pages = {210501},
	publisher = {American Physical Society},
	title = {Topological Order at Nonzero Temperature},
	url = {https://link.aps.org/doi/10.1103/PhysRevLett.107.210501},
	volume = {107},
	year = {2011}}

@article{Baskaran2008,
	author = {Baskaran, G. and Sen, Diptiman and Shankar, R.},
	doi = {10.1103/PhysRevB.78.115116},
	issue = {11},
	journal = {Phys. Rev. B},
	month = {Sep},
	numpages = {8},
	pages = {115116},
	publisher = {American Physical Society},
	title = {Spin-$S$ Kitaev model: Classical ground states, order from disorder, and exact correlation functions},
	url = {https://link.aps.org/doi/10.1103/PhysRevB.78.115116},
	volume = {78},
	year = {2008}}

@article{Peres1996,
	author = {Peres, Asher},
	doi = {10.1103/PhysRevLett.77.1413},
	issue = {8},
	journal = {Phys. Rev. Lett.},
	month = {Aug},
	numpages = {0},
	pages = {1413--1415},
	publisher = {American Physical Society},
	title = {Separability Criterion for Density Matrices},
	url = {https://link.aps.org/doi/10.1103/PhysRevLett.77.1413},
	volume = {77},
	year = {1996}}

@article{Toth2005,
	author = {T\'oth, G\'eza},
	doi = {10.1103/PhysRevA.71.010301},
	issue = {1},
	journal = {Phys. Rev. A},
	month = {Jan},
	numpages = {4},
	pages = {010301},
	publisher = {American Physical Society},
	title = {Entanglement witnesses in spin models},
	url = {https://link.aps.org/doi/10.1103/PhysRevA.71.010301},
	volume = {71},
	year = {2005}}

@article{Lake2013,
	author = {Lake, B. and Tennant, D. A. and Caux, J.-S. and Barthel, T. and Schollw\"ock, U. and Nagler, S. E. and Frost, C. D.},
	doi = {10.1103/PhysRevLett.111.137205},
	issue = {13},
	journal = {Phys. Rev. Lett.},
	month = {Sep},
	numpages = {5},
	pages = {137205},
	publisher = {American Physical Society},
	title = {Multispinon Continua at Zero and Finite Temperature in a Near-Ideal Heisenberg Chain},
	url = {https://link.aps.org/doi/10.1103/PhysRevLett.111.137205},
	volume = {111},
	year = {2013}}

@article{Shannon2012,
	author = {Shannon, Nic and Sikora, Olga and Pollmann, Frank and Penc, Karlo and Fulde, Peter},
	doi = {10.1103/PhysRevLett.108.067204},
	issue = {6},
	journal = {Phys. Rev. Lett.},
	month = {Feb},
	numpages = {5},
	pages = {067204},
	publisher = {American Physical Society},
	title = {Quantum Ice: A Quantum Monte Carlo Study},
	url = {https://link.aps.org/doi/10.1103/PhysRevLett.108.067204},
	volume = {108},
	year = {2012}}

@article{Huang2018,
	author = {Huang, Chun-Jiong and Deng, Youjin and Wan, Yuan and Meng, Zi Yang},
	doi = {10.1103/PhysRevLett.120.167202},
	issue = {16},
	journal = {Phys. Rev. Lett.},
	month = {Apr},
	numpages = {6},
	pages = {167202},
	publisher = {American Physical Society},
	title = {Dynamics of Topological Excitations in a Model Quantum Spin Ice},
	url = {https://link.aps.org/doi/10.1103/PhysRevLett.120.167202},
	volume = {120},
	year = {2018}}

@article{Kato2015,
	author = {Kato, Yasuyuki and Onoda, Shigeki},
	doi = {10.1103/PhysRevLett.115.077202},
	issue = {7},
	journal = {Phys. Rev. Lett.},
	month = {Aug},
	numpages = {5},
	pages = {077202},
	publisher = {American Physical Society},
	title = {Numerical Evidence of Quantum Melting of Spin Ice: Quantum-to-Classical Crossover},
	url = {https://link.aps.org/doi/10.1103/PhysRevLett.115.077202},
	volume = {115},
	year = {2015}}

@article{Gohlke2023,
	author = {Matthias Gohlke and Atsushi Iwaki and Chisa Hotta},
	doi = {10.21468/SciPostPhys.15.5.206},
	journal = {SciPost Phys.},
	pages = {206},
	publisher = {SciPost},
	title = {{Thermal pure matrix product state in two dimensions: Tracking thermal equilibrium from paramagnet down to the Kitaev honeycomb spin liquid state}},
	url = {https://scipost.org/10.21468/SciPostPhys.15.5.206},
	volume = {15},
	year = {2023}}

@article{Li2020,
	author = {Li, Han and Qu, Dai-Wei and Zhang, Hao-Kai and Jia, Yi-Zhen and Gong, Shou-Shu and Qi, Yang and Li, Wei},
	doi = {10.1103/PhysRevResearch.2.043015},
	issue = {4},
	journal = {Phys. Rev. Res.},
	month = {Oct},
	numpages = {16},
	pages = {043015},
	publisher = {American Physical Society},
	title = {Universal thermodynamics in the Kitaev fractional liquid},
	url = {https://link.aps.org/doi/10.1103/PhysRevResearch.2.043015},
	volume = {2},
	year = {2020}}

@article{Yamaji2016,
	author = {Yamaji, Youhei and Suzuki, Takafumi and Yamada, Takuto and Suga, Sei-ichiro and Kawashima, Naoki and Imada, Masatoshi},
	doi = {10.1103/PhysRevB.93.174425},
	issue = {17},
	journal = {Phys. Rev. B},
	month = {May},
	numpages = {14},
	pages = {174425},
	publisher = {American Physical Society},
	title = {{Clues and criteria for designing a Kitaev spin liquid revealed by thermal and spin excitations of the honeycomb iridate ${\mathrm{Na}}_{2}{\mathrm{IrO}}_{3}$}},
	url = {https://link.aps.org/doi/10.1103/PhysRevB.93.174425},
	volume = {93},
	year = {2016}}

@article{Gohlke2017,
	author = {Gohlke, Matthias and Verresen, Ruben and Moessner, Roderich and Pollmann, Frank},
	doi = {10.1103/PhysRevLett.119.157203},
	issue = {15},
	journal = {Phys. Rev. Lett.},
	month = {Oct},
	numpages = {6},
	pages = {157203},
	publisher = {American Physical Society},
	title = {Dynamics of the Kitaev-Heisenberg Model},
	url = {https://link.aps.org/doi/10.1103/PhysRevLett.119.157203},
	volume = {119},
	year = {2017}}

@article{Abanin2025,
	author = {{Google Quantum AI} and Collaborators},
	date = {2025/10/01},
	doi = {10.1038/s41586-025-09526-6},
	id = {Abanin2025},
	isbn = {1476-4687},
	journal = {Nature},
	number = {8086},
	pages = {825--830},
	title = {Observation of constructive interference at the edge of quantum ergodicity},
	url = {https://doi.org/10.1038/s41586-025-09526-6},
	volume = {646},
	year = {2025}}

@article{Gohlke2018,
	author = {Gohlke, Matthias and Wachtel, Gideon and Yamaji, Youhei and Pollmann, Frank and Kim, Yong Baek},
	doi = {10.1103/PhysRevB.97.075126},
	issue = {7},
	journal = {Phys. Rev. B},
	month = {Feb},
	numpages = {14},
	pages = {075126},
	publisher = {American Physical Society},
	title = {Quantum spin liquid signatures in Kitaev-like frustrated magnets},
	url = {https://link.aps.org/doi/10.1103/PhysRevB.97.075126},
	volume = {97},
	year = {2018}}

@article{Knolle2015,
	author = {Knolle, J. and Kovrizhin, D. L. and Chalker, J. T. and Moessner, R.},
	doi = {10.1103/PhysRevB.92.115127},
	issue = {11},
	journal = {Phys. Rev. B},
	month = {Sep},
	numpages = {20},
	pages = {115127},
	publisher = {American Physical Society},
	title = {Dynamics of fractionalization in quantum spin liquids},
	url = {https://link.aps.org/doi/10.1103/PhysRevB.92.115127},
	volume = {92},
	year = {2015}}

@article{Hermanns2018,
	author = {Hermanns, M. and Kimchi, I. and Knolle, J.},
	doi = {https://doi.org/10.1146/annurev-conmatphys-033117-053934},
	issn = {1947-5462},
	journal = {Annual Review of Condensed Matter Physics},
	number = {Volume 9, 2018},
	pages = {17-33},
	publisher = {Annual Reviews},
	title = {Physics of the Kitaev Model: Fractionalization, Dynamic Correlations, and Material Connections},
	type = {Journal Article},
	url = {https://www.annualreviews.org/content/journals/10.1146/annurev-conmatphys-033117-053934},
	volume = {9},
	year = {2018}}

@article{Rousochatzakis2019,
	author = {Rousochatzakis, I. and Kourtis, S. and Knolle, J. and Moessner, R. and Perkins, N. B.},
	doi = {10.1103/PhysRevB.100.045117},
	issue = {4},
	journal = {Phys. Rev. B},
	month = {Jul},
	numpages = {14},
	pages = {045117},
	publisher = {American Physical Society},
	title = {Quantum spin liquid at finite temperature: Proximate dynamics and persistent typicality},
	url = {https://link.aps.org/doi/10.1103/PhysRevB.100.045117},
	volume = {100},
	year = {2019}}

@article{Jiang2012,
	author = {Jiang, Hong-Chen and Wang, Zhenghan and Balents, Leon},
	date = {2012/12/01},
	doi = {10.1038/nphys2465},
	id = {Jiang2012},
	isbn = {1745-2481},
	journal = {Nature Physics},
	number = {12},
	pages = {902--905},
	title = {Identifying topological order by entanglement entropy},
	url = {https://doi.org/10.1038/nphys2465},
	volume = {8},
	year = {2012}}

@article{Laeuchli2018,
	author = {L\"auchli, Andreas M. and Sudan, Julien and Moessner, Roderich},
	doi = {10.1103/PhysRevB.100.155142},
	issue = {15},
	journal = {Phys. Rev. B},
	month = {Oct},
	numpages = {7},
	pages = {155142},
	publisher = {American Physical Society},
	title = {$S=\frac{1}{2}$ kagome Heisenberg antiferromagnet revisited},
	url = {https://link.aps.org/doi/10.1103/PhysRevB.100.155142},
	volume = {100},
	year = {2019}}

@article{Hofmann2014,
	author = {Hofmann, Martin and Moroder, Tobias and G{\"u}hne, Otfried},
	doi = {10.1088/1751-8113/47/15/155301},
	journal = {J. Phys. A: Math. Theor.},
	month = {mar},
	number = {15},
	pages = {155301},
	publisher = {IOP Publishing},
	title = {Analytical characterization of the genuine multiparticle negativity},
	url = {https://doi.org/10.1088/1751-8113/47/15/155301},
	volume = {47},
	year = {2014}}

@article{Vidal1999,
	author = {Vidal, Guifr{\'e} and Tarrach, Rolf},
	journal = {Physical Review A},
	number = {1},
	pages = {141},
	publisher = {APS},
	title = {Robustness of entanglement},
	volume = {59},
	year = {1999}}

@article{Steiner2003,
	author = {Steiner, Michael},
	journal = {Physical Review A},
	number = {5},
	pages = {054305},
	publisher = {APS},
	title = {Generalized robustness of entanglement},
	volume = {67},
	year = {2003}}

@article{Hphi-update,
	author = {Kota Ido and Mitsuaki Kawamura and Yuichi Motoyama and Kazuyoshi Yoshimi and Youhei Yamaji and Synge Todo and Naoki Kawashima and Takahiro Misawa},
	doi = {https://doi.org/10.1016/j.cpc.2024.109093},
	issn = {0010-4655},
	journal = {Computer Physics Communications},
	pages = {109093},
	title = {{Update of H$\Phi$: Newly added functions and methods in versions 2 and 3}},
	url = {https://www.sciencedirect.com/science/article/pii/S001046552400016X},
	volume = {298},
	year = {2024}}

@article{Baskaran2007,
	author = {Baskaran, G. and Mandal, Saptarshi and Shankar, R.},
	doi = {10.1103/PhysRevLett.98.247201},
	issue = {24},
	journal = {Phys. Rev. Lett.},
	month = {Jun},
	numpages = {4},
	pages = {247201},
	publisher = {American Physical Society},
	title = {Exact Results for Spin Dynamics and Fractionalization in the Kitaev Model},
	url = {https://link.aps.org/doi/10.1103/PhysRevLett.98.247201},
	volume = {98},
	year = {2007}}

@article{Nasu2015,
	author = {Nasu, Joji and Udagawa, Masafumi and Motome, Yukitoshi},
	doi = {10.1103/PhysRevB.92.115122},
	issue = {11},
	journal = {Phys. Rev. B},
	month = {Sep},
	numpages = {6},
	pages = {115122},
	publisher = {American Physical Society},
	title = {Thermal fractionalization of quantum spins in a Kitaev model: Temperature-linear specific heat and coherent transport of Majorana fermions},
	url = {https://link.aps.org/doi/10.1103/PhysRevB.92.115122},
	volume = {92},
	year = {2015}}

@article{Kawamura2017,
	author = {Kawamura, Mitsuaki and Yoshimi, Kazuyoshi and Misawa, Takahiro and Yamaji, Youhei and Todo, Synge and Kawashima, Naoki},
	journal = {Computer Physics Communications},
	pages = {180--192},
	publisher = {Elsevier},
	title = {{Quantum lattice model solver H$\Phi$}},
	volume = {217},
	year = {2017}}

@article{Wang2025,
	author = {Wang, Ting-Tung and Song, Menghan and Lyu, Liuke and Witczak-Krempa, William and Meng, Zi Yang},
	date = {2025/01/02},
	doi = {10.1038/s41467-024-55354-z},
	id = {Wang2025},
	isbn = {2041-1723},
	journal = {Nature Communications},
	number = {1},
	pages = {96},
	title = {Entanglement microscopy and tomography in many-body systems},
	url = {https://doi.org/10.1038/s41467-024-55354-z},
	volume = {16},
	year = {2025}}

@article{Yao2010,
	author = {Yao, Hong and Qi, Xiao-Liang},
	doi = {10.1103/PhysRevLett.105.080501},
	issue = {8},
	journal = {Phys. Rev. Lett.},
	month = {Aug},
	numpages = {4},
	pages = {080501},
	publisher = {American Physical Society},
	title = {Entanglement Entropy and Entanglement Spectrum of the Kitaev Model},
	url = {https://link.aps.org/doi/10.1103/PhysRevLett.105.080501},
	volume = {105},
	year = {2010}}

@article{Lecheminant1997,
	author = {Lecheminant, P. and Bernu, B. and Lhuillier, C. and Pierre, L. and Sindzingre, P.},
	doi = {10.1103/PhysRevB.56.2521},
	issue = {5},
	journal = {Phys. Rev. B},
	month = {Aug},
	numpages = {0},
	pages = {2521--2529},
	publisher = {American Physical Society},
	title = {Order versus disorder in the quantum Heisenberg antiferromagnet on the kagom\'e lattice using exact spectra analysis},
	url = {https://link.aps.org/doi/10.1103/PhysRevB.56.2521},
	volume = {56},
	year = {1997}}

@article{Norman2016,
	author = {Norman, M. R.},
	doi = {10.1103/RevModPhys.88.041002},
	issue = {4},
	journal = {Rev. Mod. Phys.},
	month = {Dec},
	numpages = {14},
	pages = {041002},
	publisher = {American Physical Society},
	title = {Colloquium: Herbertsmithite and the search for the quantum spin liquid},
	url = {https://link.aps.org/doi/10.1103/RevModPhys.88.041002},
	volume = {88},
	year = {2016}}

@article{Iqbal2011,
	author = {Iqbal, Yasir and Becca, Federico and Poilblanc, Didier},
	doi = {10.1103/PhysRevB.84.020407},
	issue = {2},
	journal = {Phys. Rev. B},
	month = {Jul},
	numpages = {4},
	pages = {020407},
	publisher = {American Physical Society},
	title = {Projected wave function study of ${\mathbb{Z}}_{2}$ spin liquids on the kagome lattice for the spin-$\frac{1}{2}$ quantum Heisenberg antiferromagnet},
	url = {https://link.aps.org/doi/10.1103/PhysRevB.84.020407},
	volume = {84},
	year = {2011}}

@misc{Matsuda-arXiv.2501,
	archiveprefix = {arXiv},
	author = {Matsuda, Yuji and Shibauchi, Takasada and Kee, Hae-Young},
	eprint = {2501.05608},
	primaryclass = {cond-mat.str-el},
	title = {Kitaev Quantum Spin Liquids},
	year = {2025}}

@article{Liao2017,
	author = {Liao, H. J. and Xie, Z. Y. and Chen, J. and Liu, Z. Y. and Xie, H. D. and Huang, R. Z. and Normand, B. and Xiang, T.},
	doi = {10.1103/PhysRevLett.118.137202},
	issue = {13},
	journal = {Phys. Rev. Lett.},
	month = {Mar},
	numpages = {6},
	pages = {137202},
	publisher = {American Physical Society},
	title = {Gapless Spin-Liquid Ground State in the $S=1/2$ Kagome Antiferromagnet},
	url = {https://link.aps.org/doi/10.1103/PhysRevLett.118.137202},
	volume = {118},
	year = {2017}}

@article{Singh2007,
	author = {Singh, Rajiv R. P. and Huse, David A.},
	doi = {10.1103/PhysRevB.76.180407},
	issue = {18},
	journal = {Phys. Rev. B},
	month = {Nov},
	numpages = {4},
	pages = {180407},
	publisher = {American Physical Society},
	title = {Ground state of the spin-1/2 kagome-lattice Heisenberg antiferromagnet},
	url = {https://link.aps.org/doi/10.1103/PhysRevB.76.180407},
	volume = {76},
	year = {2007}}

@article{Chalker1992,
	author = {Chalker, J. T. and Eastmond, J. F. G.},
	doi = {10.1103/PhysRevB.46.14201},
	issue = {21},
	journal = {Phys. Rev. B},
	month = {Dec},
	numpages = {0},
	pages = {14201--14204},
	publisher = {American Physical Society},
	title = {Ground-state disorder in the spin-1/2 kagom\'e Heisenberg antiferromagnet},
	url = {https://link.aps.org/doi/10.1103/PhysRevB.46.14201},
	volume = {46},
	year = {1992}}

@article{Hastings2000,
	author = {Hastings, M. B.},
	doi = {10.1103/PhysRevB.63.014413},
	issue = {1},
	journal = {Phys. Rev. B},
	month = {Dec},
	numpages = {16},
	pages = {014413},
	publisher = {American Physical Society},
	title = {Dirac structure, RVB, and Goldstone modes in the kagom\'e antiferromagnet},
	url = {https://link.aps.org/doi/10.1103/PhysRevB.63.014413},
	volume = {63},
	year = {2000}}

@article{Sachdev1992,
	author = {Sachdev, Subir},
	doi = {10.1103/PhysRevB.45.12377},
	issue = {21},
	journal = {Phys. Rev. B},
	month = {Jun},
	numpages = {0},
	pages = {12377--12396},
	publisher = {American Physical Society},
	title = {Kagom\'e and triangular-lattice Heisenberg antiferromagnets: Ordering from quantum fluctuations and quantum-disordered ground states with unconfined bosonic spinons},
	url = {https://link.aps.org/doi/10.1103/PhysRevB.45.12377},
	volume = {45},
	year = {1992}}

@article{He2017,
	author = {He, Yin-Chen and Zaletel, Michael P. and Oshikawa, Masaki and Pollmann, Frank},
	doi = {10.1103/PhysRevX.7.031020},
	issue = {3},
	journal = {Phys. Rev. X},
	month = {Jul},
	numpages = {16},
	pages = {031020},
	publisher = {American Physical Society},
	title = {Signatures of Dirac Cones in a DMRG Study of the Kagome Heisenberg Model},
	url = {https://link.aps.org/doi/10.1103/PhysRevX.7.031020},
	volume = {7},
	year = {2017}}

@article{Braunstein1994,
	author = {Braunstein, Samuel L. and Caves, Carlton M.},
	doi = {10.1103/PhysRevLett.72.3439},
	issue = {22},
	journal = {Phys. Rev. Lett.},
	month = {May},
	numpages = {0},
	pages = {3439--3443},
	publisher = {American Physical Society},
	title = {Statistical distance and the geometry of quantum states},
	url = {https://link.aps.org/doi/10.1103/PhysRevLett.72.3439},
	volume = {72},
	year = {1994}}

@article{Terhal2000,
	author = {Barbara M. Terhal},
	doi = {https://doi.org/10.1016/S0375-9601(00)00401-1},
	issn = {0375-9601},
	journal = {Physics Letters A},
	number = {5},
	pages = {319-326},
	title = {Bell inequalities and the separability criterion},
	url = {https://www.sciencedirect.com/science/article/pii/S0375960100004011},
	volume = {271},
	year = {2000}}

@misc{Zhou-arXiv.2510,
	archiveprefix = {arXiv},
	author = {Chengkang Zhou and Zhengbang Zhou and F{\'e}lix Desrochers and Yong Baek Kim and Zi Yang Meng},
	eprint = {2510.14813},
	primaryclass = {cond-mat.str-el},
	title = {Quantum Fisher Information as a Thermal and Dynamical Probe in Frustrated Magnets: Insights from Quantum Spin Ice},
	year = {2025}}

@article{Gao2025,
	author = {Gao, Bin and Desrochers, F{\'e}lix and Tam, David W. and Kirschbaum, Diana M. and Steffens, Paul and Hiess, Arno and Nguyen, Duy Ha and Su, Yixi and Cheong, Sang-Wook and Paschen, Silke and Kim, Yong Baek and Dai, Pengcheng},
	date = {2025/08/01},
	doi = {10.1038/s41567-025-02922-9},
	id = {Gao2025},
	isbn = {1745-2481},
	journal = {Nature Physics},
	number = {8},
	pages = {1203--1210},
	title = {Neutron scattering and thermodynamic evidence for emergent photons and fractionalization in a pyrochlore spin ice},
	url = {https://doi.org/10.1038/s41567-025-02922-9},
	volume = {21},
	year = {2025}}

@article{Lee2012,
	author = {Lee, SungBin and Onoda, Shigeki and Balents, Leon},
	doi = {10.1103/PhysRevB.86.104412},
	issue = {10},
	journal = {Phys. Rev. B},
	month = {Sep},
	numpages = {11},
	pages = {104412},
	publisher = {American Physical Society},
	title = {Generic quantum spin ice},
	url = {https://link.aps.org/doi/10.1103/PhysRevB.86.104412},
	volume = {86},
	year = {2012}}

@article{Smith2025,
	author = {Smith, Evan M. and Lhotel, Elsa and Petit, Sylvain and Gaulin, Bruce D.},
	doi = {https://doi.org/10.1146/annurev-conmatphys-041124-015101},
	issn = {1947-5462},
	journal = {Annual Review of Condensed Matter Physics},
	number = {Volume 16, 2025},
	pages = {387-415},
	publisher = {Annual Reviews},
	title = {Experimental Insights into Quantum Spin Ice Physics in Dipole--Octupole Pyrochlore Magnets},
	type = {Journal Article},
	url = {https://www.annualreviews.org/content/journals/10.1146/annurev-conmatphys-041124-015101},
	volume = {16},
	year = {2025}}

@article{Sibille2018,
	author = {Sibille, Romain and Gauthier, Nicolas and Yan, Han and Ciomaga Hatnean, Monica and Ollivier, Jacques and Winn, Barry and Filges, Uwe and Balakrishnan, Geetha and Kenzelmann, Michel and Shannon, Nic and Fennell, Tom},
	date = {2018/07/01},
	doi = {10.1038/s41567-018-0116-x},
	id = {Sibille2018},
	isbn = {1745-2481},
	journal = {Nature Physics},
	number = {7},
	pages = {711--715},
	title = {{Experimental signatures of emergent quantum electrodynamics in Pr$_2$Hf$_2$O$_7$}},
	url = {https://doi.org/10.1038/s41567-018-0116-x},
	volume = {14},
	year = {2018}}

@article{Poree2025,
	author = {Por{\'e}e, Victor and Yan, Han and Desrochers, F{\'e}lix and Petit, Sylvain and Lhotel, Elsa and Appel, Markus and Ollivier, Jacques and Kim, Yong Baek and Nevidomskyy, Andriy H. and Sibille, Romain},
	date = {2025/01/01},
	doi = {10.1038/s41567-024-02711-w},
	id = {Por{\'e}e2025},
	isbn = {1745-2481},
	journal = {Nature Physics},
	number = {1},
	pages = {83--88},
	title = {Evidence for fractional matter coupled to an emergent gauge field in a quantum spin ice},
	url = {https://doi.org/10.1038/s41567-024-02711-w},
	volume = {21},
	year = {2025}}

@article{Anderson1987,
	author = {P. W. Anderson},
	doi = {10.1126/science.235.4793.1196},
	journal = {Science},
	number = {4793},
	pages = {1196-1198},
	title = {The Resonating Valence Bond State in {${\text{La}}_{2}{\text{CuO}}_{4}$} and Superconductivity},
	url = {https://www.science.org/doi/abs/10.1126/science.235.4793.1196},
	volume = {235},
	year = {1987}}

@article{Pomeranchuck1941,
	author = {Pomeranchuck, I},
	journal = {J. Phys. 4},
	pages = {357},
	title = {The thermal conductivity of the paramagentic dielectrics at low temepratures},
	volume = {4},
	year = {1941}}

@article{Hulthen1938,
	author = {Hulth\'en, Lamek},
	journal = {Ark. Mat. Astro. Fys.},
	number = {11},
	volume = {26A},
	year = {1938}}

@article{Hams2000,
	author = {Hams, Anthony and De Raedt, Hans},
	doi = {10.1103/PhysRevE.62.4365},
	issue = {3},
	journal = {Phys. Rev. E},
	month = {Sep},
	numpages = {0},
	pages = {4365--4377},
	publisher = {American Physical Society},
	title = {{Fast algorithm for finding the eigenvalue distribution of very large matrices}},
	url = {https://link.aps.org/doi/10.1103/PhysRevE.62.4365},
	volume = {62},
	year = {2000}}

@article{Diamond2016,
	author = {Steven Diamond and Stephen Boyd},
	journal = {Journal of Machine Learning Research},
	title = {{CVXPY}: A {P}ython-Embedded Modeling Language for Convex Optimization},
	url = {https://stanford.edu/~boyd/papers/pdf/cvxpy_paper.pdf},
	year = {2016}}

@manual{MOSEK-documentation,
	author = {{MOSEK ApS}},
	title = {The MOSEK Python Fusion API manual. Version 11.0.},
	url = {https://docs.mosek.com/latest/pythonfusion/index.html},
	year = 2025}

@inproceedings{Gurvits2003,
	author = {Gurvits, Leonid},
	booktitle = {Proceedings of the thirty-fifth annual ACM symposium on Theory of computing},
	pages = {10--19},
	title = {Classical deterministic complexity of Edmonds' problem and quantum entanglement},
	year = {2003}}

@article{Wootters1998,
	author = {Wootters, William K},
	doi = {10.1098/rsta.1998.0244},
	issn = {1364503X},
	journal = {Philosophical Transactions of the Royal Society A: Mathematical, Physical and Engineering Sciences},
	number = {1743},
	pages = {1717--1731},
	title = {{Quantum entanglement as a quantifiable resource}},
	volume = {356},
	year = {1998}}

@article{Guehne2006,
	author = {G\"uhne, Otfried and T\'oth, G\'eza},
	doi = {10.1103/PhysRevA.73.052319},
	issue = {5},
	journal = {Phys. Rev. A},
	month = {May},
	numpages = {9},
	pages = {052319},
	publisher = {American Physical Society},
	title = {Energy and multipartite entanglement in multidimensional and frustrated spin models},
	url = {https://link.aps.org/doi/10.1103/PhysRevA.73.052319},
	volume = {73},
	year = {2006}}

@article{Doherty2014,
	author = {Doherty, Andrew C},
	doi = {10.1088/1751-8113/47/42/424004},
	journal = {J. Phys. A: Math. Theor.},
	month = {oct},
	number = {42},
	pages = {424004},
	publisher = {IOP Publishing},
	title = {Entanglement and the shareability of quantum states},
	url = {https://doi.org/10.1088/1751-8113/47/42/424004},
	volume = {47},
	year = {2014}}

@article{Kim2023,
	author = {Kim, Youngseok and Eddins, Andrew and Anand, Sajant and Wei, Ken Xuan and van den Berg, Ewout and Rosenblatt, Sami and Nayfeh, Hasan and Wu, Yantao and Zaletel, Michael and Temme, Kristan and Kandala, Abhinav},
	date = {2023/06/01},
	doi = {10.1038/s41586-023-06096-3},
	id = {Kim2023},
	isbn = {1476-4687},
	journal = {Nature},
	number = {7965},
	pages = {500--505},
	title = {Evidence for the utility of quantum computing before fault tolerance},
	url = {https://doi.org/10.1038/s41586-023-06096-3},
	volume = {618},
	year = {2023}}

@book{Wen2007,
	author = {Wen, Xiao-Gang},
	doi = {10.1093/acprof:oso/9780199227259.001.0001},
	isbn = {9780199227259},
	month = {09},
	publisher = {Oxford University Press},
	title = {Quantum Field Theory of Many-Body Systems: From the Origin of Sound to an Origin of Light and Electrons},
	url = {https://doi.org/10.1093/acprof:oso/9780199227259.001.0001},
	year = {2007}}

@book{Zeng2019,
	author = {Zeng, Bei and Chen, Xie and Zhou, Duan-Lu and Wen, Xiao-Gang},
	publisher = {Springer New York},
	title = {Quantum Information Meets Quantum Matter},
	year = {2019}}

@book{Skrzypczyk2023,
	author = {Skrzypczyk, Paul and Cavalcanti, Daniel},
	publisher = {IOP Publishing},
	title = {Semidefinite programming in quantum information science},
	year = {2023}}

@article{Schrodinger1935-Naturwissenschaften23,
	author = {Schr{\"o}dinger, E.},
	date = {1935/11/01},
	doi = {10.1007/BF01491891},
	id = {Schr{\"o}dinger1935},
	isbn = {1432-1904},
	journal = {Naturwissenschaften},
	number = {48},
	pages = {807--812},
	title = {Die gegenw{\"a}rtige Situation in der Quantenmechanik},
	url = {https://doi.org/10.1007/BF01491891},
	volume = {23},
	year = {1935}}

@misc{Lyu-arXiv.2505,
	archiveprefix = {arXiv},
	author = {Liuke Lyu and Deeksha Chandorkar and Samarth Kapoor and So Takei and Erik S. S{\o}rensen and William Witczak-Krempa},
	eprint = {2505.18124},
	primaryclass = {cond-mat.str-el},
	title = {Multiparty entanglement loops in quantum spin liquids},
	url = {https://arxiv.org/abs/2505.18124},
	year = {2025}}

@article{Jungnitsch2011,
	author = {Jungnitsch, Bastian and Moroder, Tobias and G\"uhne, Otfried},
	doi = {10.1103/PhysRevLett.106.190502},
	issue = {19},
	journal = {Phys. Rev. Lett.},
	month = {May},
	numpages = {4},
	pages = {190502},
	publisher = {American Physical Society},
	title = {Taming Multiparticle Entanglement},
	url = {https://link.aps.org/doi/10.1103/PhysRevLett.106.190502},
	volume = {106},
	year = {2011}}

@article{Broholm2020,
	author = {C. Broholm and R. J. Cava and S. A. Kivelson and D. G. Nocera and M. R. Norman and T. Senthil},
	doi = {10.1126/science.aay0668},
	journal = {Science},
	number = {6475},
	pages = {eaay0668},
	title = {Quantum spin liquids},
	url = {https://www.science.org/doi/abs/10.1126/science.aay0668},
	volume = {367},
	year = {2020}}

@article{Bennett1996,
	author = {Bennett, Charles H. and Bernstein, Herbert J. and Popescu, Sandu and Schumacher, Benjamin},
	doi = {10.1103/PhysRevA.53.2046},
	issue = {4},
	journal = {Phys. Rev. A},
	month = {Apr},
	numpages = {0},
	pages = {2046--2052},
	publisher = {American Physical Society},
	title = {Concentrating partial entanglement by local operations},
	url = {https://link.aps.org/doi/10.1103/PhysRevA.53.2046},
	volume = {53},
	year = {1996}}

@article{Horodecki1996,
	author = {Horodecki, Michal and Horodecki, Pawel and Horodecki, Ryszard},
	journal = {Physics Letters A},
	number = {210},
	pages = {00706--2},
	title = {Separability of mixed states: necessary and sufficient conditions},
	volume = {223},
	year = {1996}}

@article{Guehne2005,
	author = {G{\"u}hne, Otfried and T{\'o}th, G{\'e}za and Briegel, Hans J},
	journal = {New Journal of Physics},
	number = {1},
	pages = {229},
	publisher = {IOP Publishing},
	title = {Multipartite entanglement in spin chains},
	volume = {7},
	year = {2005}}

@article{Guehne2009,
	author = {G{\"u}hne, Otfried and T{\'o}th, G{\'e}za},
	journal = {Physics Reports},
	number = {1-6},
	pages = {1--75},
	publisher = {Elsevier},
	title = {Entanglement detection},
	volume = {474},
	year = {2009}}

@misc{Shimokawa-arXiv.2505,
	archiveprefix = {arXiv},
	author = {Tokuro Shimokawa and Snigdh Sabharwal and Nic Shannon},
	eprint = {2505.11874},
	primaryclass = {cond-mat.str-el},
	title = {Can experimentally-accessible measures of entanglement distinguish quantum spin liquids from disorder-driven "random singlet" phases ?},
	year = {2025}}

@article{Sabharwal2025,
	author = {Sabharwal, Snigdh and Shimokawa, Tokuro and Shannon, Nic},
	doi = {10.1103/95fl-rxl3},
	issue = {2},
	journal = {Phys. Rev. Res.},
	month = {Jun},
	numpages = {16},
	pages = {023271},
	publisher = {American Physical Society},
	title = {Witnessing disorder in quantum magnets},
	url = {https://link.aps.org/doi/10.1103/95fl-rxl3},
	volume = {7},
	year = {2025}}

@article{Schroedinger1935-MathProcCamPhilSoc31,
	author = {Schr{\"o}dinger, E.},
	doi = {10.1017/S0305004100013554},
	journal = {Mathematical Proceedings of the Cambridge Philosophical Society},
	number = {4},
	pages = {555--563},
	title = {Discussion of Probability Relations between Separated Systems},
	volume = {31},
	year = {1935}}

@article{Scheie2024,
	author = {Scheie, A. O. and Ghioldi, E. A. and Xing, J. and Paddison, J. A. M. and Sherman, N. E. and Dupont, M. and Sanjeewa, L. D. and Lee, Sangyun and Woods, A. J. and Abernathy, D. and Pajerowski, D. M. and Williams, T. J. and Zhang, Shang-Shun and Manuel, L. O. and Trumper, A. E. and Pemmaraju, C. D. and Sefat, A. S. and Parker, D. S. and Devereaux, T. P. and Movshovich, R. and Moore, J. E. and Batista, C. D. and Tennant, D. A.},
	date = {2024/01/01},
	doi = {10.1038/s41567-023-02259-1},
	id = {Scheie2024},
	isbn = {1745-2481},
	journal = {Nature Physics},
	number = {1},
	pages = {74--81},
	title = {Proximate spin liquid and fractionalization in the triangular antiferromagnet {KYbSe2}},
	volume = {20},
	year = {2024}}

@article{Laurell2024,
	author = {Laurell, Pontus and Scheie, Allen and Dagotto, Elbio and Tennant, D. Alan},
	doi = {https://doi.org/10.1002/qute.202400196},
	journal = {Advanced Quantum Technologies},
	number = {n/a},
	pages = {2400196},
	title = {Witnessing Entanglement and Quantum Correlations in Condensed Matter: A Review},
	url = {https://onlinelibrary.wiley.com/doi/abs/10.1002/qute.202400196},
	volume = {8},
	year = {2024}}

@article{Hauke2016,
	author = {Hauke, Philipp and Heyl, Markus and Tagliacozzo, Luca and Zoller, Peter},
	date = {2016/08/01},
	doi = {10.1038/nphys3700},
	id = {Hauke2016},
	isbn = {1745-2481},
	journal = {Nature Physics},
	number = {8},
	pages = {778--782},
	title = {Measuring multipartite entanglement through dynamic susceptibilities},
	url = {https://doi.org/10.1038/nphys3700},
	volume = {12},
	year = {2016}}

@article{Sugiura2013,
	author = {Sugiura, Sho and Shimizu, Akira},
	doi = {10.1103/PhysRevLett.111.010401},
	issue = {1},
	journal = {Phys. Rev. Lett.},
	month = {Jul},
	numpages = {5},
	pages = {010401},
	publisher = {American Physical Society},
	title = {Canonical Thermal Pure Quantum State},
	url = {https://link.aps.org/doi/10.1103/PhysRevLett.111.010401},
	volume = {111},
	year = {2013}}

@article{Imada1986,
	author = {Imada ,Masatoshi and Takahashi ,Minoru},
	doi = {10.1143/JPSJ.55.3354},
	eprint = {https://doi.org/10.1143/JPSJ.55.3354},
	journal = {Journal of the Physical Society of Japan},
	number = {10},
	pages = {3354-3361},
	title = {{Quantum Transfer Monte Carlo Method for Finite Temperature Properties and Quantum Molecular Dynamics Method for Dynamical Correlation Functions}},
	url = {https://doi.org/10.1143/JPSJ.55.3354},
	volume = {55},
	year = {1986}}

@article{Scheie2021,
	author = {Scheie, A. and Laurell, Pontus and Samarakoon, A. M. and Lake, B. and Nagler, S. E. and Granroth, G. E. and Okamoto, S. and Alvarez, G. and Tennant, D. A.},
	doi = {10.1103/PhysRevB.103.224434},
	issue = {22},
	journal = {Phys. Rev. B},
	month = {Jun},
	numpages = {16},
	pages = {224434},
	publisher = {American Physical Society},
	title = {Witnessing entanglement in quantum magnets using neutron scattering},
	url = {https://link.aps.org/doi/10.1103/PhysRevB.103.224434},
	volume = {103},
	year = {2021}}

@article{Wu2019,
	author = {Wu, Han-Qing and Gong, Shou-Shu and Sheng, D. N.},
	doi = {10.1103/PhysRevB.99.085141},
	issue = {8},
	journal = {Phys. Rev. B},
	month = {Feb},
	numpages = {13},
	pages = {085141},
	publisher = {American Physical Society},
	title = {Randomness-induced spin-liquid-like phase in the spin-$\frac{1}{2}$ ${J}_{1}\ensuremath{-}{J}_{2}$ triangular Heisenberg model},
	url = {https://link.aps.org/doi/10.1103/PhysRevB.99.085141},
	volume = {99},
	year = {2019}}

@article{Laflorancie2016,
	author = {Nicolas Laflorencie},
	doi = {https://doi.org/10.1016/j.physrep.2016.06.008},
	issn = {0370-1573},
	journal = {Physics Reports},
	note = {Quantum entanglement in condensed matter systems},
	pages = {1-59},
	title = {Quantum entanglement in condensed matter systems},
	url = {https://www.sciencedirect.com/science/article/pii/S0370157316301582},
	volume = {646},
	year = {2016}}

@article{Levin2006,
	author = {Levin, Michael and Wen, Xiao-Gang},
	doi = {10.1103/PhysRevLett.96.110405},
	issue = {11},
	journal = {Phys. Rev. Lett.},
	month = {Mar},
	numpages = {4},
	pages = {110405},
	publisher = {American Physical Society},
	title = {Detecting Topological Order in a Ground State Wave Function},
	url = {https://link.aps.org/doi/10.1103/PhysRevLett.96.110405},
	volume = {96},
	year = {2006}}

@article{Kitaev2006-PRL96,
	author = {Kitaev, Alexei and Preskill, John},
	doi = {{10.1103/PhysRevLett.96.110404}},
	issue = {11},
	journal = {Phys. Rev. Lett.},
	month = {Mar},
	numpages = {4},
	pages = {110404},
	publisher = {American Physical Society},
	title = {{Topological Entanglement Entropy}},
	url = {{https://link.aps.org/doi/10.1103/PhysRevLett.96.110404}},
	volume = {96},
	year = {2006}}

@article{Kitaev2006-AnnPhys321,
	author = {Alexei Kitaev},
	doi = {http://dx.doi.org/10.1016/j.aop.2005.10.005},
	issn = {0003-4916},
	journal = {Annals of Physics},
	note = {{January Special Issue}},
	number = {1},
	pages = {2 - 111},
	title = {{Anyons in an exactly solved model and beyond}},
	url = {http://www.sciencedirect.com/science/article/pii/S0003491605002381},
	volume = {321},
	year = {2006}}

@article{Savary2016,
	author = {Lucile Savary and Leon Balents},
	doi = {10.1088/0034-4885/80/1/016502},
	journal = {Reports on Progress in Physics},
	month = {nov},
	number = {1},
	pages = {016502},
	publisher = {{IOP} Publishing},
	title = {Quantum spin liquids: a review},
	url = {https://doi.org/10.1088/0034-4885/80/1/016502},
	volume = {80},
	year = 2016}

@article{Lee2008,
	author = {Lee, Patrick A.},
	doi = {10.1126/science.1163196},
	issn = {0036-8075},
	journal = {Science},
	number = {5894},
	pages = {1306--1307},
	publisher = {American Association for the Advancement of Science},
	title = {An End to the Drought of Quantum Spin Liquids},
	url = {http://science.sciencemag.org/content/321/5894/1306},
	volume = {321},
	year = {2008}}

@article{Balents2010,
	author = {Balents, Leon},
	issn = {0028-0836},
	journal = {Nature},
	month = {mar},
	number = {7286},
	pages = {199--208},
	publisher = {Nature Publishing Group},
	title = {{Spin liquids in frustrated magnets}},
	url = {http://dx.doi.org/10.1038/nature08917},
	volume = {464},
	year = {2010}}

@article{Anderson1973,
	author = {P.W. Anderson},
	doi = {http://dx.doi.org/10.1016/0025-5408(73)90167-0},
	issn = {0025-5408},
	journal = {Materials Research Bulletin},
	number = {2},
	pages = {153 - 160},
	title = {Resonating valence bonds: A new kind of insulator?},
	url = {http://www.sciencedirect.com/science/article/pii/0025540873901670},
	volume = {8},
	year = {1973}}

@article{Machida2012,
	author = {Machida, Manabu and Iitaka, Toshiaki and Miyashita, Seiji},
	doi = {10.1103/PhysRevB.86.224412},
	issue = {22},
	journal = {Phys. Rev. B},
	month = {Dec},
	numpages = {8},
	pages = {224412},
	publisher = {American Physical Society},
	title = {{ESR intensity and the Dzyaloshinsky-Moriya interaction of the nanoscale molecular magnet V${}_{15}$}},
	url = {https://link.aps.org/doi/10.1103/PhysRevB.86.224412},
	volume = {86},
	year = {2012}}

@article{Iitaka2003,
	author = {Iitaka, Toshiaki and Ebisuzaki, Toshikazu},
	doi = {10.1103/PhysRevLett.90.047203},
	issue = {4},
	journal = {Phys. Rev. Lett.},
	month = {Jan},
	numpages = {4},
	pages = {047203},
	publisher = {American Physical Society},
	title = {{Algorithm for Linear Response Functions at Finite Temperatures: Application to ESR Spectrum of $s=\frac{1}{2}$ Antiferromagnet Cu Benzoate}},
	url = {https://link.aps.org/doi/10.1103/PhysRevLett.90.047203},
	volume = {90},
	year = {2003}}

@article{Toth2012,
	author = {T\'oth, G\'eza},
	doi = {10.1103/PhysRevA.85.022322},
	issue = {2},
	journal = {Phys. Rev. A},
	month = {Feb},
	numpages = {8},
	pages = {022322},
	publisher = {American Physical Society},
	title = {Multipartite entanglement and high-precision metrology},
	url = {https://link.aps.org/doi/10.1103/PhysRevA.85.022322},
	volume = {85},
	year = {2012}}

@article{Hyllus2012,
	author = {Hyllus, Philipp and Laskowski, Wies\l{}aw and Krischek, Roland and Schwemmer, Christian and Wieczorek, Witlef and Weinfurter, Harald and Pezz\'e, Luca and Smerzi, Augusto},
	doi = {10.1103/PhysRevA.85.022321},
	issue = {2},
	journal = {Phys. Rev. A},
	month = {Feb},
	numpages = {10},
	pages = {022321},
	publisher = {American Physical Society},
	title = {Fisher information and multiparticle entanglement},
	url = {https://link.aps.org/doi/10.1103/PhysRevA.85.022321},
	volume = {85},
	year = {2012}}

@article{Jiang2023,
	author = {Jiang, Yi-Fan and Jiang, Hong-Chen},
	doi = {10.1103/PhysRevB.107.L140411},
	issue = {14},
	journal = {Phys. Rev. B},
	month = {Apr},
	numpages = {5},
	pages = {L140411},
	publisher = {American Physical Society},
	title = {Nature of quantum spin liquids of the $S=\frac{1}{2}$ Heisenberg antiferromagnet on the triangular lattice: A parallel DMRG study},
	url = {https://link.aps.org/doi/10.1103/PhysRevB.107.L140411},
	volume = {107},
	year = {2023}}

@article{Endo2018,
	author = {Endo, Hiroyuki and Hotta, Chisa and Shimizu, Akira},
	doi = {10.1103/PhysRevLett.121.220601},
	issue = {22},
	journal = {Phys. Rev. Lett.},
	month = {Nov},
	numpages = {6},
	pages = {220601},
	publisher = {American Physical Society},
	title = {From Linear to Nonlinear Responses of Thermal Pure Quantum States},
	url = {https://link.aps.org/doi/10.1103/PhysRevLett.121.220601},
	volume = {121},
	year = {2018}}

@article{Ikeuchi2015,
	author = {Ikeuchi, Hiroki and De Raedt, Hans and Bertaina, Sylvain and Miyashita, Seiji},
	doi = {10.1103/PhysRevB.92.214431},
	issue = {21},
	journal = {Phys. Rev. B},
	month = {Dec},
	numpages = {15},
	pages = {214431},
	publisher = {American Physical Society},
	title = {{Computation of ESR spectra from the time evolution of the magnetization: Comparison of autocorrelation and Wiener-Khinchin-relation-based methods}},
	url = {https://link.aps.org/doi/10.1103/PhysRevB.92.214431},
	volume = {92},
	year = {2015}}

@article{Benton2012,
	author = {Benton, Owen and Sikora, Olga and Shannon, Nic},
	journal = {Phys. Rev. B},
	pages = {075154},
	title = {{Seeing the light: Experimental signatures of emergent electromagnetism in a quantum spin ice}},
	url = {https://journals.aps.org/prb/abstract/10.1103/PhysRevB.86.075154},
	volume = {86},
	year = {2012}}

@article{Banerjee2008,
	author = {Banerjee, Argha and Isakov, Sergei V. and Damle, Kedar and Kim, Yong Baek},
	journal = {Phys. Rev. Lett.},
	optdoi = {10.1103/PhysRevLett.100.047208},
	optissue = {4},
	optmonth = {Jan},
	optnumpages = {4},
	optpublisher = {American Physical Society},
	opturl = {http://link.aps.org/doi/10.1103/PhysRevLett.100.047208},
	pages = {047208},
	title = {Unusual Liquid State of Hard-Core Bosons on the Pyrochlore Lattice},
	url = {http://link.aps.org/doi/10.1103/PhysRevLett.100.047208},
	volume = {100},
	year = {2008}}

@article{Hermele2004,
	author = {Hermele, Michael and Fisher, Matthew P. A. and Balents, Leon},
	journal = {Phys. Rev. B},
	optdoi = {10.1103/PhysRevB.69.064404},
	optissue = {6},
	optmonth = {Feb},
	optnumpages = {21},
	optpublisher = {American Physical Society},
	opturl = {http://link.aps.org/doi/10.1103/PhysRevB.69.064404},
	pages = {064404},
	title = {Pyrochlore photons: The \protect{U(1)} spin liquid in a \protect{S=$\frac{1}{2}$} three-dimensional frustrated magnet},
	url = {http://link.aps.org/doi/10.1103/PhysRevB.69.064404},
	volume = {69},
	year = {2004}}

@article{Banerjee2016,
	author = {Banerjee, A. and Bridges, C. A. and Yan, J. -Q. and Aczel, A. A. and Li, L. and Stone, M. B. and Granroth, G. E. and Lumsden, M. D. and Yiu, Y. and Knolle, J. and Bhattacharjee, S. and Kovrizhin, D. L. and Moessner, R. and Tennant, D. A. and Mandrus, D. G. and Nagler, S. E.},
	date = {2016/07//print},
	isbn = {1476-1122},
	journal = {Nat. Mater.},
	l3 = {10.1038/nmat4604; http://www.nature.com/nmat/journal/v15/n7/abs/nmat4604.html#supplementary-information},
	m3 = {Article},
	month = {07},
	number = {7},
	pages = {733--740},
	publisher = {Nature Publishing Group},
	title = {Proximate Kitaev quantum spin liquid behaviour in a honeycomb magnet},
	ty = {JOUR},
	url = {http://dx.doi.org/10.1038/nmat4604},
	volume = {15},
	year = {2016}}

@article{Jackeli2009,
	author = {Jackeli, G. and Khaliullin, G.},
	doi = {10.1103/PhysRevLett.102.017205},
	issue = {1},
	journal = {Phys. Rev. Lett.},
	month = {Jan},
	numpages = {4},
	pages = {017205},
	publisher = {American Physical Society},
	title = {Mott Insulators in the Strong Spin-Orbit Coupling Limit: From Heisenberg to a Quantum Compass and Kitaev Models},
	url = {https://link.aps.org/doi/10.1103/PhysRevLett.102.017205},
	volume = {102},
	year = {2009}}

@article{Shimokawa2016,
	author = {Shimokawa ,Tokuro and Kawamura ,Hikaru},
	doi = {10.7566/JPSJ.85.113702},
	eprint = {https://doi.org/10.7566/JPSJ.85.113702},
	journal = {Journal of the Physical Society of Japan},
	number = {11},
	pages = {113702},
	title = {Finite-Temperature Crossover Phenomenon in the S = 1/2 Antiferromagnetic Heisenberg Model on the Kagome Lattice},
	url = {https://doi.org/10.7566/JPSJ.85.113702},
	volume = {85},
	year = {2016}}

@article{Yan2011,
	author = {Yan, Simeng and Huse, David A. and White, Steven R.},
	doi = {10.1126/science.1201080},
	issn = {0036-8075},
	journal = {Science},
	number = {6034},
	pages = {1173--1176},
	publisher = {American Association for the Advancement of Science},
	title = {Spin-Liquid Ground State of the S = 1/2 Kagome Heisenberg Antiferromagnet},
	url = {https://science.sciencemag.org/content/332/6034/1173},
	volume = {332},
	year = {2011}}

@article{Sorensen2001,
	author = {S\o{}rensen, Anders S. and M\o{}lmer, Klaus},
	doi = {10.1103/PhysRevLett.86.4431},
	issue = {20},
	journal = {Phys. Rev. Lett.},
	month = {May},
	numpages = {0},
	pages = {4431--4434},
	publisher = {American Physical Society},
	title = {Entanglement and Extreme Spin Squeezing},
	url = {https://link.aps.org/doi/10.1103/PhysRevLett.86.4431},
	volume = {86},
	year = {2001}}

@misc{Datta2009,
	archiveprefix = {arXiv},
	author = {Nilanjana Datta},
	eprint = {0807.2536},
	primaryclass = {quant-ph},
	title = {Max- Relative Entropy of Entanglement, alias Log Robustness},
	year = {2009}}
%%%%%%%%%%%%%%%%%%%%%%%%%%%%%%%%%%%%%%

%%%%%%%%%%%%%%%%%%%%%%%%%%%%%%%%%%%%%%
% End matter
%%%%%%%%%%%%%%%%%%%%%%%%%%%%%%%%%%%%%%

 \balancecolsandclearpage

%%%%%%%%%%%%%%%%%%%%%%%%%%%%%%%%%%%%%%%%%
\appendix				
%%%%%%%%%%%%%%%%%%%%%%%%%%%%%%%%%%%%%%%%%

\section*{End Matter}
\label{section:end.matter}

\begin{figure}[t]
    \centering
    	\subfloat[ Kagome AF \label{fig:kagome.finite.size} ]{\includegraphics[height=0.4\linewidth]{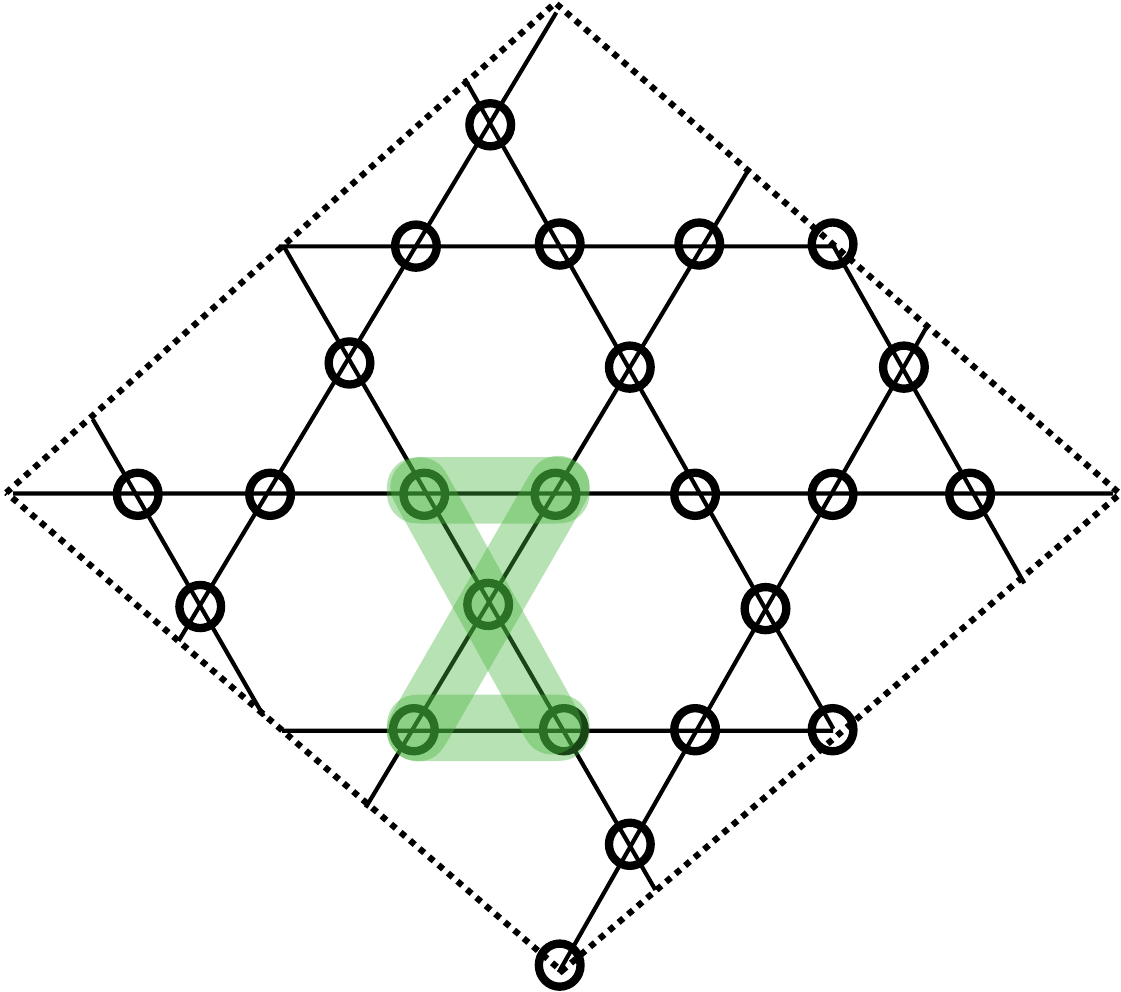}}
	\hspace{0.5cm}
	\subfloat[ Kitaev model \label{fig:honeycomb.finite.size} ]{\includegraphics[height=0.4\linewidth]{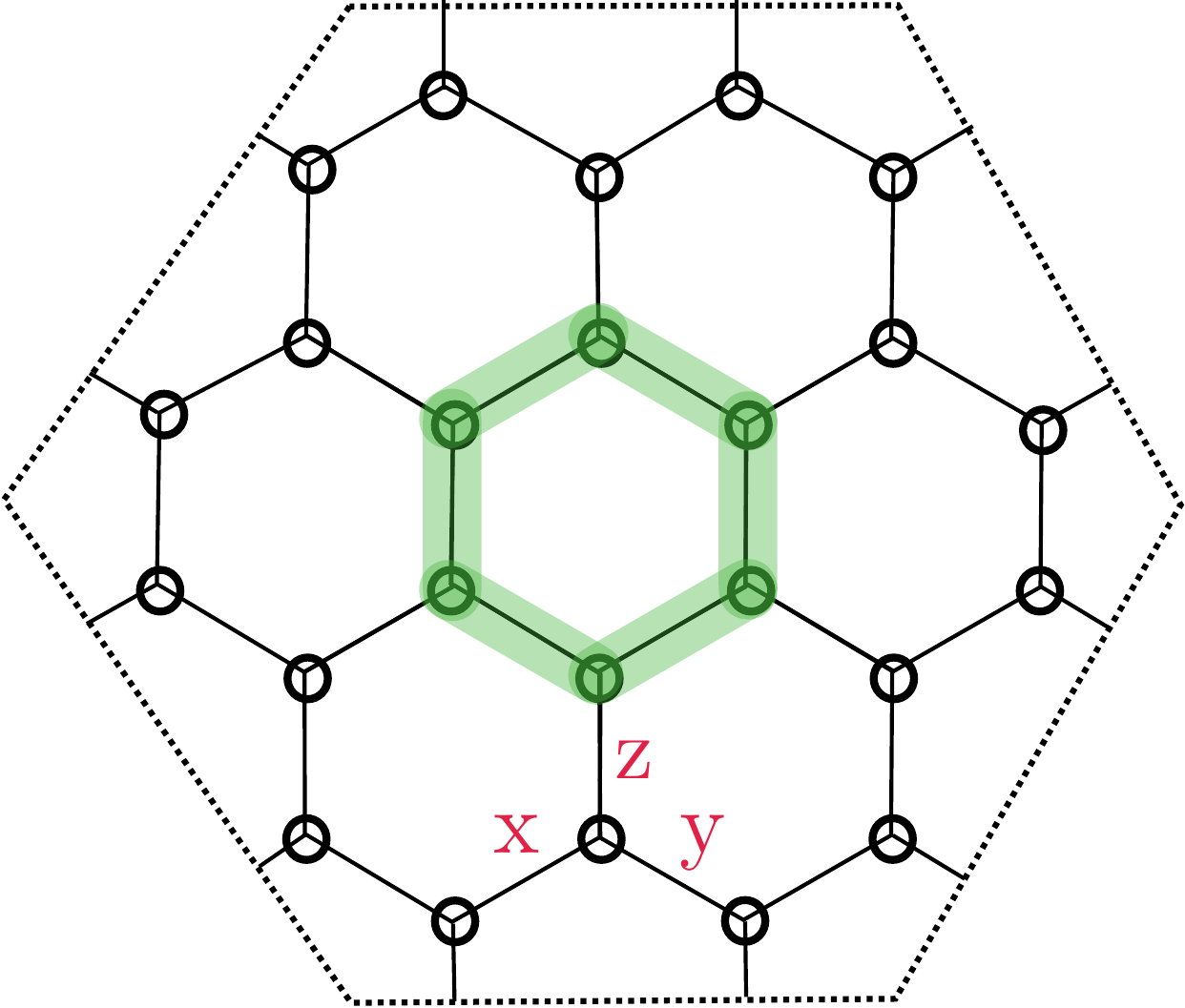}}
    \caption{Clusters of the used in calculations for the Kagome lattice antiferromagnet (AF) 
    and the Kitaev model on a honeycomb lattice.   
    (a) 24--site cluster on Kagome lattice, showing location of 6-bond cycle used to calculate results 
    for entanglement given in the main text.
    (b) 24--site cluster with full symmetry of honeycomb lattice, showing location of 6--bond cycle 
    considered in the main text.
    In both cases, periodic boundary conditions are imposed, shown here through the 
    dashed boundary of the cluster.
    }
    \label{fig:clusters-for-calculation}
\end{figure}

%%%%%%%%%%%%%%%%%%%%%%%%%%%%%%%%%%%%%%%%%%%%%%%%%%%%%%%%%%%%%%%%%%%%%%%
{\it Models.} 
%%%%%%%%%%%%%%%%%%%%%%%%%%%%%%%%%%%%%%%%%%%%%%%%%%%%%%%%%%%%%%%%%%%%%%%
The first model we consider is the spin--1/2 Heisenberg antiferromagnet 
on a Kagome lattice (KAF)
\begin{eqnarray}
{\mathcal H}_{\sf KAF} 
	&=& J \sum_{\langle ij \rangle} {\bf S}_i \cdot {\bf S}_j  \quad  \quad [J > 0] \; , 
\label{eq:H.Kagome}
\end{eqnarray}
where the sum $\langle ij \rangle$ runs over the 1$^{st}$--neighbour bonds of 
a Kagome lattice.
The calculations described in the main text were carried out for  a 24--site cluster 
with periodic boundary conditions, illustrated in Fig.~\ref{fig:kagome.finite.size}.

%%%%%%%%%%%%%%%%%%%%%%%%%%%%%%%%%%%%%%

The second model we consider is the Kitaev model on honeycomb lattice 
(KHM) \cite{Kitaev2006-AnnPhys321} 
\begin{eqnarray}
{\mathcal H}_{\sf KHM} 
	&=&    \sum_\alpha \sum_{\langle ij \rangle \in \{\alpha\}} J_\alpha S_i^\alpha  S_j^\alpha  \; , 
\label{eq:H.Kitaev}
\end{eqnarray}
where the sum on $\alpha = \{x,y,x\}$ runs over the three inequivalent 
1$^{st}$--neighbour bonds of a honeycomb lattice, 
and we consider the isotropic parameter set 
\begin{eqnarray}
	  J_\alpha \equiv J > 0 \; . 
\end{eqnarray}
The calculations described in the main text were carried out for a 24--site cluster 
with periodic boundary conditions, illustrated in Fig.~\ref{fig:honeycomb.finite.size}.  

%%%%%%%%%%%%%%%%%%%%%%%%%%%%%%%%%%%%%%

In the case of the Kitaev model, it is also useful to consider the operator 
\begin{eqnarray}
\mathcal{W}_p
	&=& \sigma_1^x  \sigma_2^y \sigma_3^z \sigma_4^x  \sigma_5^y \sigma_6^z\; , 
\label{eq:W_p}
\end{eqnarray}
which measures the $\mathbb{Z}_2$ gauge flux 
on an hexagonal plaquette of the lattice \cite{Kitaev2006-AnnPhys321}.
Within the Kitaev QSL, in the limit $T \to 0$,  $\langle \mathcal{W}_p \rangle \to  1$.    
\\

%%%%%%%%%%%%%%%%%%%%%%%%%%%%%%%%%%%%%%%%%%%%%%%%%%%%%%%%%%%%%%%%%%%%%%%
{\it Primal and Dual SDPs for entanglement depth.}
%%%%%%%%%%%%%%%%%%%%%%%%%%%%%%%%%%%%%%%%%%%%%%%%%%%%%%%%%%%%%%%%%%%%%%%
Recall from the main text that we denote by $\mathscr{P}_{n, k-1}$ the set of partitions (subsystems) of $n$ qubits to depth $k-1$, i.e. with each subsystem containing at most $k-1$ qubits. We denote an element of this set as $\pi$, 
and a subsystem within $\pi$ as $\mathcal{s}$. To each  partition we associate a state $\rho_\pi$, and demand that $\rho_\pi^{T_{\mathcal{s}}}\succeq 0$ for each $\mathcal{s}$.

%%%%%%%%%%%%%%%%%%%%%%%%%%%%%%%%%%%%%%%%%%%%%%%%%%%%%%%%%%%%%%%%%%%%%%%

If $\rho_n(T)$ can be written as a mixture $\rho_n(T) = \sum_\pi q_\pi \rho_\pi$, with $q_\pi$ probabilities, (satisfying $q_\pi \geq 0$ $\forall \pi$ and $\sum_\pi q_{\pi} = 1$) then it is said to be $(k-1)$-producible. If $\rho_n(T)$ cannot be decomposed this way, we say it has entanglement to (at least) depth $k$. 

%%%%%%%%%%%%%%%%%%%%%%%%%%%%%%%%%%%%%%%%%%%%%%%%%%%%%%%%%%%%%%%%%%%%%%%

We can further make this quantitative. If a state $\rho_n(T)$ has depth-$k$ entanglement, we can quantify \textit{how much} depth-$k$ entanglement it has by considering the amount of ``noise'' we must mix with it in order for it to become $(k-1)$-producible. That is, we can seek to find the minimum value of $r$ such that the mixture 
\begin{equation}\label{e:k-1-prod}
\frac{\rho_n(T) + r\sigma}{1+r}
\end{equation}
is $(k-1)$-producible, where $\sigma$ is an arbitrary density matrix (possible highly entangled) and the mixing parameter is written as $1/(1+r)$, with $0 \leq r \leq \infty$. Such a parameterisation both allows for this to be cast as an SDP (see below), and for the optimal $r^*$, the quantity  $\ln (1+r^*)$ can be interpreted as a type of generalized relative entropy \cite{Datta2009}.

%%%%%%%%%%%%%%%%%%%%%%%%%%%%%%%%%%%%%%%%%%%%%%%%%%%%%%%%%%%%%%%%%%%%%%%
%  Fig. 4 - figure defining clusters used in simulation
%%%%%%%%%%%%%%%%%%%%%%%%%%%%%%%%%%%%%%%%%%%%%%%%%%%%%%%%%%%%%%%%%%%%%%%

%%%%%%%%%%%%%%%%%%%%%%%%%%%%%%%%%%%%%%%%%%%%%%%%%%%%%%%%%%%%%%%%%%%%%%%

To turn this into an SDP, we define new variables $\tilde{\rho}_\pi = (1+r)q_\pi \rho_\pi \succeq 0$ and $\tilde\sigma = r\sigma \succeq 0$, which we can view as unnormalised density matrices. In terms of these variables, \eqref{e:k-1-prod} is $(k-1)$-producible if $\rho_n(T) + \tilde{\sigma} = \sum_\pi \tilde\rho_\pi$. 
Taking the trace of both sides of this we see that $1 + r = \Tr \sum_\pi \tilde\rho_\pi$. We can then (i) solve for $\tilde{\sigma}$, i.e.~$\tilde\sigma = \sum_\pi \tilde\rho_\pi - \rho_n(T) \succeq 0$ and (ii) note that $\rho_\pi^{T_\mathcal{s}} \succeq 0$ if and only if $\tilde\rho_\pi^{T_\mathcal{s}} \succeq 0$ (since positive rescaling don't affect the positivity of eigenvalues of matrices). Putting everything together, minimising $(1+r)$ is equivalent to the SDP optimization problem stated in the main text, namely 
\begin{align}\label{eq:depth.of.entanglement-SDP-appendix}
	\min_{\{\tilde{\rho}_\pi\}} & \quad   \Tr \hspace{-1em}\sum_{\pi \in \mathscr{P}_{n,k-1}} \tilde{\rho}_\pi \\
	\text{s.t.}	 &   \quad	\sum_{\pi \in \mathscr{P}_{n,k-1}} \tilde{\rho}_\pi  \succeq \rho_n(T), \nonumber \\
    & \quad \tilde{\rho}_\pi \succeq 0 \quad \forall \pi \in \mathscr{P}_{n,k-1}, \nonumber \\
    & \quad \tilde{\rho}_\pi^{T_\mathcal{s}} \succeq 0 \quad \forall \mathcal{s} \in \pi, \forall \pi \in \mathscr{P}_{n,k-1}, \nonumber 
\end{align}
where we recall that $A \succeq B$ is shorthand for $A - B \succeq 0$. This is now manifestly an SDP, with a linear objective function, and all constraints being linear inequality constraints. 

Applying the duality theory of SDPs, we can obtain a second (equivalent) optimization problem, which also evaluates to $1 + \mathcal{R}_k[\rho_n(T)]$. In particular, by introducing appropriate Lagrange multipliers for each constraint, forming a Lagrangian, and making it independent of the (primal) variables, we arrive at the following dual SDP:
\begin{align}\label{e: dual SDP depth}
	\max_{W,\{Y_\pi\}} & \quad   \Tr [W \rho_n(T)] \\
	\text{s.t.}	 &   \quad	\openone - \sum_{\mathcal{s} \in \pi} Y_\pi^{T_\mathcal{s}} \succeq W \quad \forall \pi \in \mathscr{P}_{n,k-1} \nonumber \\
    & \quad W \succeq 0, \quad Y_\pi^\mathcal{s} \succeq 0 \quad \forall \mathcal{s} \in \pi, \forall \pi \in \mathscr{P}_{n,k-1}. \nonumber
\end{align}
Eq.~\eqref{e: dual SDP depth}, is an optimization over quantitative entanglement witnesses for depth-$k$ entanglement: The constraints of the dual SDP ensure that $W$ is a quantitative witness for depth-$k$ entanglement. The expectation value of $W$ evaluated on any state will be non-zero only if the state has depth-$k$ entanglement, and the value will always provide a lower bound on $1 + \mathcal{R}_k [\rho_n(T)]$. The SDP optimizes over all such witnesses, and is guaranteed to find a witness $W^*$ such that $1 + \mathcal{R}_k [\rho_n(T)] = \Tr[W^*\rho_n(T)]$. 

If we consider the case of full depth, $k=n$, we are testing whether or not a state can be written as a mixture of states, each of which can be entangled between up to $n-1$ qubits (i.e.~is $(n-1)$-producible). In this case, if $(n-1)$ qubits can be entangled, there is only 1 remaining qubit, and hence we consider all \textit{bipartitions} of the state. 
This is precisely what is considered in the context of \textit{genuine multipartite entanglement} (GME), discussed below.

%%%%%%%%%%%%%%%%%%%%%%%%%%%%%%%%%%%%%%%%%%%%%%%%%%%%%%%%%%%%%%%%%%%%%%%
{\it Geometric structure of entanglement. }
%%%%%%%%%%%%%%%%%%%%%%%%%%%%%%%%%%%%%%%%%%%%%%%%%%%%%%%%%%%%%%%%%%%%%%%
Entanglement depth is agnostic about the underlying geometric structure of the subsystems considered. It only takes into account the maxmium number of qubits (or particles) within a given subsystem. When considering specific physical systems, not all partitions/subsystems are equal, and hence it can be insightful also to consider specific set of partitions, and ask whether a state is `producible' without entanglement, relative to this given set.   
Answering this question sheds light on the geometrical structure of the (multipartite) 
entanglement in the system, and can be accomplished by restricting the set of partitions considered
in the SDP for entanglement depth to the reduced set ${\mathcal P}_\lambda$ [Eq.~(\ref{eq:S.restricted.partitions})].
As a specific example, we consider the 6-bond cycle of the Kiteav model, and label the 6 qubits by $ABCDEF$ in clockwise fashion, starting at an arbitrary qubit, with $A$ and $F$ neighbours [cf. inset to Fig.~\ref{fig:results-for-Kitaev-model}(b)]

A state is said to be genuine multipartite entangled (GME) if it cannot be written as a convex 
mixture of states which are separable with respect to any bipartition \cite{Guehne2009}. 
In its simplest form, GME is defined for three qubits $A$, $B$ and $C$, 
and expressed in terms of the possible bipartitions of these qubits, 
\begin{eqnarray}
 	\mathscr{P}_{\sf ABC} = \{ A|BC, B|AC, C|AB \} \; .
\end{eqnarray}
The simplest way to generalize to a larger number of qubits is to divide these into three 
distinct ``parties'', and then to proceed by analogy with the 
example of three qubits.
This was the approach taken in \cite{Lyu-arXiv.2505}, where the authors considered the set 
\begin{eqnarray}
    \mathscr{P}^{(3)}_{\hexagon} = \{ AB|CDEF, CD|ABEF, EF|ABCD\} 
\label{eq:delta.P.3}
\end{eqnarray}
This can be interpreted as looking at genuine tripartite entanglement between 3 subsystems, $AB$, $CD$, $EF$, each comprising two qubits -- a qubit and its neighbour on one side. 

More generally, however, GME can be defined through any set of bipartitions which spans the 
qubits under considersation.  
Since the specifc choice $\mathscr{P}^{(3)}_{\hexagon}$ [Eq.~(\ref{eq:delta.P.3})]
breaks a symmetry of the cycle, this motivates us to consider an alternative set of 6 partitions,  
\begin{multline}
\mathscr{P}_{\hexagon}^{\sf (2)} = \{ AB|CDEF, CD|ABEF, EF|ABCD, \\ FA|BCDE, BC|DEFA, DE|FABC\},
\end{multline}
which restore that symmetry by pairing each qubit with both of its nearest neighbours.  
Testing for producibility relative to this geometrically-motivated set of partitions doesn't have an interpretation as GME, but can be viewed as a testing for a restricted type of depth-5 entanglement, where the 4 qubits must be contiguous on the hexagon.

The strongest measure of GME between six qubits is associated with the following partition
\begin{multline}
    \mathscr{P}_{\hexagon}^{\sf (1)} =
    \{\,A|BCDEF, B|ACDEF, C|ABDEF, \\
     D|ABCEF, E|ABCDF, F|ABCDE, \\
     AB|CDEF, AC|BDEF, AD|BCEF, AE|BCDF,  \\
     AF|BCDE, BC|ADEF, BD|ACEF, BE|ACDF, \\
     BF|ACDE, CD|ABEF, CE|ABDF, CF|ABDE, \\
     DE|ABCF,DF|ABCE, EF|ABCD, \\
     ABC|DEF, ABD|CEF, ABE|CDF, ABF|CDE,   \\
     ACD|BEF, ACE|BDF, ACF|BDE, ADE|BCF, \\
      ADF|BCE, AEF|BCD\}.
\end{multline}
Examples of partitions drawn from the sets 
$\mathscr{P}_{\hexagon}^{\sf (1)}$,
$\mathscr{P}_{\hexagon}^{\sf (2)}$,
and $\mathscr{P}_{\hexagon}^{\sf (3)}$
are shown in Fig.~\ref{fig:results-for-Kitaev-model}(b).

%%%%%%%%%%%%%%%%%%%%%%%%%%%%%%%%%%%%%%%%%%%%%%%%%%%%%%%%%%%%%%%%%%%%%%%
{\it Numerical methods.}
%%%%%%%%%%%%%%%%%%%%%%%%%%%%%%%%%%%%%%%%%%%%%%%%%%%%%%%%%%%%%%%%%%%%%%%
The reduced density matrices $\rho_n(T)$, used for calculations of entanglement, 
were evaluated for a cluster of $n$ spins on a graph embedded within a finite size cluster.
The specific cyclic graphs used to produce the results shown in Fig.~\ref{fig:results-for-Kagome} 
and Fig.~\ref{fig:results-for-Kitaev-model} of the main text are illustrated 
in Fig.~\ref{fig:clusters-for-calculation}.
Density matrices at finite temperature were evaluated using the method of Thermal Pure Quantum 
states (TPQ) \cite{Imada1986,Hams2000,Iitaka2003,Machida2012,Sugiura2013,Ikeuchi2015,Endo2018}, 
implemented through the package H$\Phi$ \cite{Kawamura2017,Hphi-update}.
At each temperature, an average 
\begin{eqnarray}
\rho_n (T) &=& \frac{1}{N_{\sf TPQ}} \sum_{i=1}^{N_{\sf TPQ}} \rho_n^{(i)} (T)
\end{eqnarray}
was taken over reduced density matrices $\rho_n^{(i)} (T)$ constructed from 
\mbox{$N_{\sf TPQ} = 24$} different realization a of TPQ state.
We have confirmed that results for GME measures are not significantly different when evaluated 
for $N_{\sf TPQ} = 50,100$.
Conventional Lanczos methods were used to evaluate reduced density matrices within the 
ground state, \mbox{$\rho_n(T=0)$}, for comparison with \cite{Lyu-arXiv.2505}, and validation of finite--temperature results.   

The expectation value of the plaquette operator $\mathcal{W}_p$~[Eq.~(\ref{eq:W_p})], 
was evaluated as 
\begin{equation}
	\langle \mathcal{W}_p \rangle = \Tr[ \rho_n  \mathcal{W}_p ] \;,
\end{equation}
Heat capacity per spin was evaluated as
\begin{eqnarray}
	c(T) &=&  \frac{\overline{\langle H^{2} \rangle} - \overline{\langle H \rangle}^2}{N T^2} \; ,
\label{eq:C.of.T}
\end{eqnarray}
where $\overline{\langle\,\cdot\,\rangle}$ corresponds to the TPQ average defined in \cite{Sugiura2013}.

Evaluation of SDP was carried out within the python convex optimization 
library CVXPY~\cite{Diamond2016}, using the MOSEK solver \cite{MOSEK-documentation} \textcolor{black}{whose implementation can be found in the github repository 
\cite{Snigdhgit2025}}

%%%%%%%%%%%%%%%%%%%%%%%%%%%%%%%%%%%%%%%%%%%%%%%%%%%%
\end{document}